\documentclass[12pt,tightenlines,eqsecnum,floats,aps,amsmath,amssymb,superscriptaddress,nofootinbib,prd,showpacs]{revtex4}

\usepackage[dvips]{graphicx}
\usepackage[mathscr]{eucal}
\usepackage{amsmath}
\usepackage{amsfonts}
\usepackage{amssymb}
\usepackage{pictex}
\usepackage{epsfig}
\usepackage{colordvi}
\usepackage{color}

\def\be{\nopagebreak[3]\begin{equation}}
\def\ee{\end{equation}}
\def\ba{\nopagebreak[3]\begin{eqnarray}}
\def\ea{\end{eqnarray}}
\def\bas{\nopagebreak[3]\begin{eqnarray*}}
\def\eas{\end{eqnarray*}}

\def\d{{\rm d}}
\def\w{\omega}

\newcommand{\teta}{\rlap{\lower2ex\hbox{$\,\tilde{}$}}\eta{}}

\newcommand{\bi}{\begin{itemize}}
\newcommand{\ei}{\end{itemize}}
\newcommand{\mc}[1]{\mathcal{#1}}

\def\lp{{\ell}_{\rm Pl}}

\def\w{{}^o\!\omega}

\def\la{\langle}
\def\ra{\rangle}

\newcommand{\f}{\frac}

\usepackage{enumerate}

\usepackage{colordvi}
\newcounter{mnotecount}[section]

\def\f{\frac}

\begin{document}
\preprint{\vbox{\baselineskip=12pt \rightline{IGC-10/12-3}
}}

\title{Loop quantum cosmology of Bianchi IX: Effective dynamics}

\author{Alejandro Corichi}\email{corichi@matmor.unam.mx}
\affiliation{Centro de Ciencias Matem\'aticas,
Universidad Nacional Aut\'onoma de M\'exico,
UNAM-Campus Morelia, A. Postal 61-3, Morelia, Michoac\'an 58090,
Mexico}
\affiliation{Center for Fundamental Theory, Institute for Gravitation and the Cosmos,
Pennsylvania State University, University Park
PA 16802, USA}

\author{Edison Montoya}
\email{emontoya@uis.edu.co}
\affiliation{Escuela de F\'{\i}sica, Universidad Industrial de Santander, 
A.A. 678, Bucaramanga, Colombia}


\begin{abstract}
\noindent
We study solutions to the effective equations for the Bianchi IX class of spacetimes
within loop quantum cosmology (LQC). We consider Bianchi IX models whose matter content
is a massless scalar field, by numerically solving the loop quantum cosmology effective equations,
with and without inverse triad corrections. 
The solutions are classified using certain geometrically motivated classical observables. 
We show that both effective theories
--with lapse $N$=$V$ and $N$=1-- resolve the big bang singularity and reproduce
the classical dynamics far from the bounce. Moreover, due to the positive spatial curvature, 
there is an infinite number of bounces and recollapses. 
We study the limit of large field momentum and show that both effective theories reproduce 
the same dynamics, thus recovering general relativity.
We implement a procedure to identify amongst the Bianchi IX solutions, those that behave like 
$k$=$0,1$ FLRW as well as Bianchi I, II, and VII${}_0$ models. The effective solutions exhibit Bianchi I
phases with Bianchi II transitions and also Bianchi VII${}_0$ phases, which had not been
studied before. We comment on the possible implications of these
results for a quantum modification to the classical BKL behaviour.
\end{abstract}

\pacs{04.60.Pp, 04.60.Bc, 98.80.Qc}
\maketitle
\section{Introduction}

In recent years various homogeneous cosmological models have been studied within the context of loop quantum 
cosmology (LQC) \cite{lqc}. In particular, for the models that have been exactly solved at the quantum level, 
the dynamics of sharply peaked, semiclassical states is very well described by an {\it effective} theory that incorporates the main quantum corrections to the dynamics \cite{aps2,slqc,CM-1,singhnew}. These examples give one confidence that the effective LQC description will provide reliable information about semiclassical states for models which have not been fully solved. It is in this context that we study the Bianchi IX model.
Within homogeneous and anisotropic models of the universe, the Bianchi IX model exhibits very interesting features. One of them is due
to the positive curvature of the underlying spatial manifold, which has the consequence that all classical solutions exhibit a recollapse for
generic matter content following Einstein's dynamics. At the effective
level that we shall consider here, an infinite number of bounces and recollapses appear, a feature that
was first seen in the Friedman-Lema\^itre-Robertson-Walker (FLRW) 
model with $k=1$ \cite{closed,CK,CK-2,ST,DS}. Furthermore, in the Bianchi IX model, the so called inverse triad
corrections of LQC play an important role and modify the quantum theory yielding different descriptions. 
A valid question is whether there exist physical criteria to select one theory over another,
as was done in the isotropic case \cite{cs:unique,sv,singh-BI}. 
In other words,
can the semiclassical limit choose the `physically correct' theory?
After all, the semiclassical limit of the quantum theory is expected to make contact with the classical
theory, and this can be useful to select the theory that satisfies some physically motivated requirements. 
If we assume that the semiclassical limit of each quantum theory is well described
by their effective theory, we can try to answer this question by studying the effective theory.
One should note that this issue has already been explored for 
some models \cite{singh-Ed,singh-flat}. Here
we want to explore this issue further, and consider an improved quantization of the Bianchi IX model, 
put forward in \cite{CK-3}, that incorporates the so-called inverse triad corrections.

The answer that we find in the numerical exploration of the different effective theories, is that each of them describes different dynamics but, as we show, in the limit of large field momentum the dynamics of all the different theories is almost the same. For the Bianchi IX models, large field momentum amounts to spacetimes that attain a large volume (in Planck units) before re-collapsing.
From this point of view, even if there are different quantum theories with different semiclassical
limits, if the universe grows to be `large' then both theories describe the same 
physics in the large field momentum regime.

The Bianchi IX model has long been studied in the context of Loop Quantum Cosmology. The first study 
\cite{Mb}, was within the so called $\mu_0$ dynamics, whose infrared limit has been shown to be unphysical
\cite{aps2,cs:unique}. In order to solve the problems with 
the infrared limit, the improved $\bar\mu$ dynamics 
for Bianchi IX  was constructed \cite{bianchiIX}. It is within the framework of the improved $\bar\mu$
dynamics that we consider the effective theories for this work. We will show how to
recover the correct infrared limit and resolve the big bang singularity in the ultraviolet limit.

Another important feature of the Bianchi IX model comes from the 
BKL (Belinskii-Khalatnikov-Lifshitz) conjecture \cite{BKL1,BKL2,BKL-AHS}, 
which suggest that the local dynamics of inhomogeneous cosmologies near the 
big bang singularity is dominated by the time derivatives, which 
become more important than the spatial derivatives. This fact 
makes the local dynamics at each spatial point independent and, therefore, can be approximated
by a homogeneous universe. In general, the dynamics is described by a 
Bianchi IX universe, which exhibits an oscillatory behaviour between
different Bianchi I phases with Bianchi II transitions.
Another important feature of the BKL conjecture and the evolution
near the big bang singularity is that, in general, the dynamical contribution 
from matter is negligible and the universe can be described as a 
vacuum universe. If one puts together these two elements that follow from
the BKL conjecture, then one can conclude that the dynamics of any universe
near the initial singularity can be described locally as a homogeneous
universe where all the dynamical contribution comes from the anisotropies
or, in other words, a vacuum Bianchi IX universe. An important caveat is that 
this is true only if the matter content is {\it different} from a massless scalar field. 
In this particular case the dynamical contribution from the matter turns out to be important
\cite{jacobs,wain_ellis}. 
This peculiar and unique behaviour of the massless scalar field justifies its study within 
the LQC Bianchi IX model. 
First, if the model under consideration consists of a scalar field together with another 
kind of matter, it is only the mass-less scalar field that 
contributes to the dynamics in a significant way. Second, if there is no scalar field
in the universe, then the conjecture states that the dynamics is dominated by a vacuum universe,
which can be seen itself as a limiting case of a universe with a massless scalar field, when the 
momentum of the field goes to zero (for more details on the classical behaviour of
Bianchi IX  see \cite{wain_ellis,Misner}). As this discussion suggests, the massless scalar field
together with the vacuum case, represent the most interesting cases to be studied in a Bianchi IX model.
In this manuscript we restrict ourselves to the massless scalar case within LQC. We shall only briefly comment on the vacuum case. 
\medskip

Since the Bianchi I and II models play an important role in the classical 
behaviour of Bianchi IX, we shall also consider the effective models of Bianchi type I and II, 
highlighting the most important aspects of these models,
with the purpose of understanding the Bianchi IX effective dynamics. Additionally,
given that the quantization process can produce unexpected results, we include also
the study of the isotropic FLRW models with $k=0,1$, in order to illustrate some
new results in the dynamics of Bianchi IX. Finally, since our study is numeric, 
we use some known results of the isotropic models and Bianchi I \cite{bianchiold,bianchiI}, 
in order to test the validity  of the numerical implementation. It is important to point out that 
these cosmological models are not sub-cases  within the quantum Bianchi IX model \cite{bianchiIX}, 
but are limiting cases instead. A special case is given by the classical isotropic  $k=1$ model
which {\it is} embedded in the Bianchi IX universe. 
For the LQC effective models, this is also the case only when the inverse triad 
corrections are {\it not} included. If one includes them, then the isotropic model is only a 
limiting case \cite{CK,CK-2,CK-3}. Here we shall show 
that in, the large momentum regime, the isotropic limits of the effective Bianchi IX 
and the effective FLRW $k=1$ coincide, regardless of whether there are inverse
triad corrections or not.
\medskip

This paper is the second in a series. In the first one \cite{CK-3}, different version of the
loop quantization of Bianchi IX model are introduced. In contrast to previous treatments, the inverse 
corrections are fully implemented. The effective theory is constructed and some of its properties 
are analysed. In the third manuscript of the series \cite{CKM-short} we shall present some discussion 
regarding qualitative features of the effective dynamics for generic Bianchi IX LQC models, 
including the vacuum case.
One should also note that a summary of some of these results has already been reported in \cite{CKM}. 
\medskip

This article is organized as follows: In Sec.~\ref{sec:2} we review the relevant results
for the classical Bianchi IX and discuss some classical observables relevant for the effective
theory. Next, in Sec.~\ref{sec:3} we summarize the two effective theories used in our study. 
In Sec.~\ref{sec:4} we discuss the numerical results of the effective theories. Since we do not have an analytical understanding of the space of solutions, nor of the ``generic" behaviour, 
we explore several limiting cases. 
Finally, the conclusions are outlined in Sec.~\ref{sec:5}. Additionally, we include an Appendix with all
the Hamiltonians and equations of motion used in this manuscript, and a second Appendix where we display the numerical convergence tests of our integrations.

\section{Classical Dynamics}
\label{sec:2}

In this section we shall recall the classical description of the Bianchi IX
model. This section has two parts. In the first one we review the basic variables
used in both the geometric and the connection dynamics formulations. In the second part,
we introduce the geometrical invariants and observables that are useful to extract physical
information.

\subsection{The model}

In the Bianchi IX model (like in the FLRW with $k=1$) the spatial manifold
 $\Sigma$ has the topology of a three sphere $\mathbb{S}^3$. The physical
 metric on $\Sigma$ is $q_{ab}:= \omega_a^i\omega_{b}^j\delta_{ij}$, 
 where the physical forms are
$\omega_a^i = a^i(t)\w_a^i$ and the fiducial ones are \cite{bianchiIX}
\bas
\w_a^1 &=& \sin\beta\sin\gamma (\d\alpha)_a + \cos\gamma (\d\beta)_a\,,  \\ 
\w_a^2 &=& -\sin\beta\cos\gamma (\d\alpha)_a + \sin\gamma (\d\beta)_a\,, \\ 
\w_a^3 &=& \cos\beta (\d\alpha)_a + (\d\gamma)_a\,,
\eas
%
with $\alpha \in [0, 2\pi )$, $\beta \in [ 0,\pi )$ and $\gamma \in [0,4\pi )$.
The radius of the three sphere as measured by the fiducial metric is 
$a_0$, therefore the volume with respect to
the fiducial metric is $V_0=2\pi^2\,a_0^3$. We define 
$\ell_0:=V_0^{1/3}$, that can be rewritten as $\ell_0=:\vartheta\, a_0$, 
with $\vartheta:=(2\pi^2)^{1/3}$. The triads and connections are  
$$
A_a^i=\f{c^i}{\ell_0}\,{}^o\!\omega^i_a\,, \quad 
E^a_i=\f{p_i}{\ell^{2}_0}\sqrt{{}^o\!q}\,{}^o\!e^a_i .
$$
Using these definitions and the lapse function 
$N=V=\sqrt{p_1p_2p_3}$ (with $p_i>0$), 
the Hamiltonian constraint is \cite{bianchiIX}
\bas   
\mathcal{C}_{H_{\rm BIX}}^{\rm Cla} 
&=& -\f{1}{8\pi G\gamma^2} \Bigg[p_1c_1p_2c_2+p_2c_2p_3 c_3+p_3c_3p_1c_1 
+ 2\vartheta \big(p_1p_2c_3 +p_2p_3c_1+p_3p_1c_2\big) \nonumber \\
& & \qquad+\vartheta^2 (1+ \gamma^2)\,\bigg[ 2p_1^2 +2p_2^2 +2p_3^2
-\Big(\f{p_2p_3}{p_1}\Big)^2 -\Big(\f{p_3p_1}{p_2}\Big)^2 
-\Big(\f{p_1p_2}{p_3}\Big)^2 \bigg]\Bigg] \nonumber \\
& & + \f{1}{2}p_\phi^2 \approx 0.
\eas 
Poisson brackets of the basic variables are $\{c^i,p_j\}=8\pi G\gamma\delta_j^i$ and $ \{\phi,p_\phi\}=1$.
The equations of motion come from the Poisson bracket of the basic variables with
the Hamiltonian constraint
\bas 
\dot{p}_i=\{p_i,\mc{C}_{H_{BIX}}\} \,,&\quad& \dot{c}^i =\{c^i,\mc{C}_{H_{BIX}}\}\,, \\
\dot{\phi}=\{\phi,\mc{C}_{H_{BIX}}\} \,,&\quad& \dot{p}_\phi=\{p_\phi,\mc{C}_{H_{BIX}}\} \,,
\eas   
which are given by
\bas 
\dot{p}_\phi& = &0 \,,\\
\dot{\phi}& = & p_\phi \,, \\
\dot{p_1}&=&\f{p_1}{\gamma}\left[p_2c_2 + p_3c_3
+ 2\vartheta \f{p_2p_3}{p_1}\right], \\
\dot{c_1} &=&- \f{1}{\gamma} \Bigg[p_2c_1c_2 + p_3c_1c_3 
+ 2\vartheta(p_2c_3 + p_3c_2) \\
& & \quad\quad + 4\vartheta^2 (1+\gamma^2) 
\bigg(p_1 + \f{p_2^2p_3^2}{2p_1^3} - \f{p_1p_3^2}{2p_2^2} - \f{p_1p_2^2}{2p_3^2}
\bigg) \Bigg]. 
\eas 
The other equations can be obtained by permutations of the labels.

\subsection{Observables}
\label{sec:observables}

In order to study the homogeneous cosmological models, there are two interesting physical time variables. The first
one is  harmonic time $\tau$, which corresponds to the choice 
of lapse equal to the volume, $N=V$. The other one is the cosmic (proper) time $t$ (corresponding to the
choice of lapse $N=1$).
These two time variables are related by the equation
\be  \label{times_rel} 
\f{\d}{\d t}= \f{1}{\sqrt{p_1p_2p_3}} \, \, \f{\d}{\d\tau}\, .
\ee
 The observables are usually defined with respect to the cosmic time. The derivative with respect to the
cosmic time is denoted as $O'={\d O}/{\d t}$, therefore $O'=\dot O/\sqrt{p_1p_2p_3}$,
where the `dot' stands for derivative with respect to harmonic time $\tau$ 
(we are using the notation from references \cite{bianchiIX,bianchiII,CM-bianchi2}).
In order to understand the singularity resolution and the evolution of the system as defined by
the classical and effective equations, we study the following observables:

\begin{enumerate}
\item {\it Directional scale factors},
$$a_i=\f{1}{\ell_0}\sqrt{\f{p_jp_k}{p_i}},$$ 
which come from the relation $p_i=a_j a_k \ell_0^2$, with {\small $i \neq j\neq k\neq i $}
and $p_i,a_i>0$.

\item {\it Directional Hubble parameters},
$$H_i = \f{a_i'}{a_i} =
\f{1}{2}\left(\f{p_j'}{p_j}+\f{p_k'}{p_k}-\f{p_i'}{p_i}\right),$$
with {\small $i \neq j\neq k\neq i $}. 

\item {\it Expansion} $$\theta=\f{V'}{V}=H_1+H_2+H_3.$$ 
An equivalent function is the 
{\it Average Hubble parameter} $$H=\f{\theta}{3}=\f{1}{3}(H_1+H_2+H_3).$$

\item {\it Matter density} $$\rho=\f{p_\phi^2}{2V^2}=\f{p_\phi^2}{2p_1p_2p_3}.$$ 
The dynamical contribution due to the matter is measured by the {\it Density parameter}
\be
\Omega :=\f{ 8\pi G}{3}\f{\rho}{H^2}\, .
\ee

\item {\it Shear} 
\footnote{Note that this definition of $\sigma^2$ differs from the standard definition $\sigma^2=\f{1}{2}\sigma_{ab}\sigma^{ab}$.}
$$\sigma^2=\sigma_{ab}\sigma^{ab}=\f{1}{3}[(H_1-H_2)^2+(H_1-H_3)^2+(H_2-H_3)^2] = \sum_{i=1}^3 H_i^2 -\f{1}{3}\theta^2\, .$$ 
The dynamical contribution due to the anisotropies is measured by the {\it Shear parameter}
\be
\Sigma^2 := \f{3\sigma^2}{2\theta^2}=\f{\sigma^2}{6H^2} .
\ee

\item {\it Ricci scalar}
\be
R=\left(\f{{p}_1'}{p_1}\right)^2+\left(\f{{p}_2'}{p_2}\right)^2
+\left(\f{{p}_3'}{p_3}\right)^2 +2\theta'\, .
\ee
The $p_i'$ and  $\theta'$ are computed from the equations of motion. In the classical case
 $\theta'$ is given by the  Raychaudhuri equation $\theta' = -\f{1}{2}\theta^2-\sigma^2-16\pi G \rho\,.$

\item {\it Intrinsic curvature}, 
one feature of Bianchi II and IX models is that the spatial curvature is different from zero.
The intrinsic spatial curvature is given by
\be \label{Ricci3}
{}^{(3)}R = -\f{1}{2}\big[x_1^2 + x_2^2 + x_3^2 - 2(x_1x_2 + x_1x_3+ x_2x_3)\big]\,,
\ee 

where
\be 
x_1 = \alpha_1\sqrt{\f{p_2p_3}{p_1^3}}\,, \quad
x_2 = \alpha_2\sqrt{\f{p_1p_3}{p_2^3}}\,, \quad
x_3 = \alpha_3\sqrt{\f{p_1p_2}{p_3^3}}\,. \label{xi}
\ee 
The values of $\alpha_i$ are: 
\bi 
\item Bianchi I, $\alpha_1=\alpha_2 =\alpha_3=0 $.
\item Bianchi II, $\alpha_1=1, \alpha_2 =0, \alpha_3=0 $ or permutations.
\item Bianchi IX, $\alpha_1=\alpha_2 =\alpha_3=l_0^2 $, with
$l_0=V_0^{1/3}$.
\item The isotropic FLRW model with $k=1$, is obtained from the Bianchi IX model by setting  $x_1=x_2=x_3$.
\ei

One can also define another quantity that gives information about the dynamical contribution of the
intrinsic curvature, named {\it Curvature parameter},
\be
K = -\f{3\, {}^{(3)}R}{2\theta^2} 
  = \f{3}{4\theta^2} \big[x_1^2 + x_2^2 + x_3^2 - 2(x_1x_2 + x_1x_3+ x_2x_3)\big]\,,
\ee

The parameters $\Omega, \Sigma^2$ and $K$ satisfied the classical relation
\be
\Omega + \Sigma^2 + K =1 \, .
\ee
As we shall see later, these parameters are ill defined at the quantum bounce (where $\theta=0$).
Therefore they are only useful away from the bounce, and in the classical region.

\item {\it Kasner exponents} 
\be \label{kasner}
k_i=\f{H_i}{|\theta|}.
\ee
These parameters are useful to determine when the solutions are of type Bianchi I.
The sign of $H_i$ determines whether the $a_i$ direction is expanding ($k_i>0$) or contracting ($k_i<0$).

\end{enumerate}


The classical Bianchi IX model does not have  complete exact solutions, so
there are only some general results. We summarize those results that
are important to us (a more complete review can be found in \cite{wain_ellis}):

\bi 
\item All the Bianchi IX solutions recollapse if the matter 
satisfies the dominant energy condition and has a non-negative average pressure 
\cite{Lin-Wald}.

\item  Bianchi IX has, as a limiting case, the Bianchi I model when
$x_i\rightarrow 0$, with $i=1,2,3$.

\item  Bianchi IX has, as a limiting case, the Bianchi II model when
$x_1\ne 0, x_2\rightarrow 0, x_3\rightarrow 0$, and permutations of $x_1, x_2, x_3$.

\item Bianchi IX reduces to $k=1$ FLRW when 
$p_1=p_2=p_3$ and $c_1=c_2=c_3$.

\item Bianchi IX has, as a limiting case, the $k=0$  FLRW  when  
$p_1=p_2=p_3$, $c_1=c_2=c_3$ and 
$x_i\rightarrow 0$, with $i=1,2,3$.

\item The solutions to the vacuum Bianchi IX can be approximated 
near the big bang singularity as evolving in Bianchi I phases
with Bianchi II transitions \cite{Misner}.

\item Bianchi IX reduces to Bianchi VII$_0$ 
when $\alpha_1=0 ,\alpha_2 =1, \alpha_3=1 $ (and permutations) and is a limiting 
case when $\alpha_1=\alpha_2 =\alpha_3=l_0^2 $, and one of the $x_i$ goes to zero,
with $i=1,2,3$.
\ei

The $x_i$ variables will be very useful when we study the evolution of the effective
solutions of Bianchi IX, because they give information about the transitions 
of the solutions. In order to distinguish between the transitions that Bianchi IX has,
we chose Bianchi IX ($\alpha_1=\alpha_2 =\alpha_3=l_0^2$) and study the evolution of
the $x_i$ variables. The different limits that the Bianchi IX solutions can approach may be 
classified as follows,
\bi 
\item Bianchi I: $(x_1\rightarrow 0, x_2\rightarrow 0, x_3\rightarrow 0)$.
\item Bianchi II: $(x_1\rightarrow 0, x_2\rightarrow 0)$ or $(x_1\rightarrow 0, x_3\rightarrow 0)$ or 
($x_2\rightarrow 0, x_3\rightarrow 0$).
\item Bianchi VII$_0$: $(x_1\rightarrow 0)$ or ($x_2\rightarrow 0$) or ($x_3\rightarrow 0$).
\ei

Note that Bianchi VII$_0$ is included in this classification given that, classically, it
forms part of the Bianchi IX dynamics. At the effective level it is not known whether this is 
still true, therefore it is good to include this case in our study. It is important to emphasize
that, so far, there does not exist a Loop Quantum Cosmology version of Bianchi VII$_0$ 
and therefore neither its effective version. Then, when we describe a particular regime behaving 
as Bianchi VII$_0$, this refers to solutions in which one of the $x_i$ goes to zero.
It would probably be worthwhile to quantize the Bianchi VII$_0$ model, obtain its effective dynamics,
and compare with what we get here for Bianchi IX and call Bianchi VII$_0$.
\smallskip

The evolution of each $x_i$ can give us information on 
how close to other Bianchi models is the solution of the full
Bianchi IX dynamics. Moreover, the values of $\Omega, \Sigma^2$ and $K$
can give us information about how the dynamics of
the system is affected due to the contributions from matter, anisotropies 
and curvature. Classically, the identity $\Omega + \Sigma^2 + K =1$ is always satisfied,
and this can be used to verify when the evolution of the system is approaching the classical region.
Furthermore, the Kasner exponents $k_i$ can be helpful to classify which kind of Bianchi I solutions
we get.

All of these elements will be useful to study the solutions to the effective equations.
Even when our goal is to study the Bianchi IX dynamics, due to its complexity it is necessary to study 
the simpler Bianchi models as well as the isotropic models. The effective Hamiltonian 
of each of these theories are shown in the next section. All the equations of motion
are displayed in Appendix~\ref{app:b3}.

\section{Effective Dynamics}
\label{sec:3}

The ``improved dynamics'' quantization of the Bianchi IX model \cite{bianchiIX}
consists in calculating the connection from the holonomies and use this connection to
define the curvature, which is then promoted to a quantum operator. There exists a freedom in 
the choice of the lapse function; this freedom is not important physically at classical level, 
but at quantum level it defines different quantum theories. In this manuscript
we consider two effective theories that come from two different quantum theories for the Bianchi IX
model. 
The first choice is when the lapse function is equal to the volume $N=V$, and
the second one when $N=1$. The effective Hamiltonian for $N=V$ is \cite{bianchiIX}
\begin{align} 
\mathcal{C}_{\rm H_{BIX}}^{(1)} =&
-\f{p_1p_2p_3}{8\pi G\gamma^2\lambda^2}
\big(\sin\bar\mu_1c_1\sin\bar\mu_2c_2+\sin\bar\mu_2c_2
\sin\bar\mu_3c_3+\sin\bar\mu_3c_3\sin\bar\mu_1c_1\big) \nonumber
\\ &  - \f{\vartheta}{4\pi G\gamma^2\lambda}\bigg(
\f{(p_1p_2)^{3/2}\!\!\!}{\sqrt{p_3}}\sin\bar\mu_3c_3+
\f{(p_2p_3)^{3/2}\!\!\!}{\sqrt{p_1}}\sin\bar\mu_1c_1+
\f{(p_3p_1)^{3/2}\!\!\!}{\sqrt{p_2}}\sin\bar\mu_2c_2\bigg) \nonumber \\
&  -\f{\vartheta^2 (1+\gamma^2)}{8\pi G \gamma^2}\bigg[2(p_1^2+p_2^2+p_3^2) -
\left(\f{p_1p_2}{p_3}\right)^2
-\left(\f{p_2p_3}{p_1}\right)^2
-\left(\f{p_3p_1}{p_2}\right)^2
\bigg] \nonumber \\
& + \f{p_\phi^2}{2} \approx 0 \,, \label{H-effec-BIX2}
\end{align}
with $\vartheta=(2\pi^2)^{1/3}$,
$\lambda^2 =4 \sqrt{3} \pi \gamma \lp^2$, $\gamma$ the Barbero-Immirzi parameter and
\[
\bar\mu_1 = \lambda\sqrt{\f{p_1}{p_2 p_3}} , \qquad
\bar\mu_2 = \lambda\sqrt{\f{p_2}{p_1 p_3}} , \qquad
\bar\mu_3 = \lambda\sqrt{\f{p_3}{p_1 p_2}} .
\]
When the lapse function is $N=1$ the effective Hamiltonian is given by \cite{Asieh,CK-3}
\ba
\mc{C}_{\rm H_{BIX}}^{(2)}&=&
-\frac{V^4A(V)h^6(V)}{8\pi GV_c^6\gamma^2\lambda^{2}}
\big(\sin\bar\mu_1c_1\sin\bar\mu_2c_2+\sin\bar\mu_1c_1\sin\bar\mu_3c_3
+\sin\bar\mu_2c_2\sin\bar\mu_3c_3\big)\nonumber\\
& &-\frac{\vartheta A(V)h^4(V)}{4\pi GV_c^4\gamma^2\lambda}\bigg(p_1^2p_2^2\sin\bar\mu_3c_3
+p_2^2p_3^2\sin\bar\mu_1c_1 +p_1^2p_3^2\sin\bar\mu_2c_2\bigg)\nonumber\\
& &-\frac{\vartheta^2(1+\gamma^2)A(V)h^4(V)}{8\pi GV_c^4\gamma^2}\times \nonumber\\
& &\quad\quad \bigg(2V[p_1^2+p_2^2+p_3^2] -\bigg[(p_1p_2)^{4}+(p_1p_3)^{4}+(p_2p_3)^{4}\bigg]\frac{h^6(V)}{V_c^6}\bigg)\nonumber\\
& &+\f{h^6(V)V^2}{2V^6_c}p_\phi^2 \approx 0\,, \label{H-effec-BIX4}
\ea
with $V_c=2\pi\gamma\lambda\lp^2$ and
\ba
h(V)&=&\sqrt{V+V_c}-\sqrt{|V-V_c|} \nonumber \,,\\
A(V)&=&\frac{1}{2V_c}(V+V_c-|V-V_c|) \nonumber\,.
\ea
The Poisson brackets are
$\{c^i,p_j\}=8\pi G\gamma\delta_j^i$ and $ \{\phi,p_\phi\}=1$. The equations of motion
are shown in appendix \ref{app:b3}.

Given that the classical Bianchi IX model has phases of Bianchi I and transitions of Bianchi II \cite{Misner},
it is interesting to study what happens at the effective level. Furthermore, we also consider the 
effective Hamiltonians for Bianchi I \cite{bianchi1} and Bianchi II \cite{bianchiII}, 
which can be written as \cite{bianchiII}
\ba
\mc{C}_{\rm H_{BII}} & = &\f{p_1p_2p_3}{8\pi G\gamma^2\lambda^2}
\left[\f{}{}\sin\bar\mu_1c_1\sin\bar\mu_2c_2+\sin\bar\mu_2c_2
\sin\bar\mu_3c_3+\sin\bar\mu_3c_3\sin\bar\mu_1c_1\right] \nonumber\\
& &\quad + \f{1}{8\pi G\gamma^2}
\Bigg[\f{\alpha(p_2p_3)^{3/2}}{\lambda\sqrt{p_1}}\sin\bar\mu_1c_1
-(1+\gamma^2)\left(\f{\alpha p_2p_3}{2p_1}\right)^2 \Bigg] - \f{p_\phi^2}{2} \approx 0 \, ,
\label{H-BII}
\ea 

where $\alpha$ plays the role of a switch between Bianchi I
($\alpha=0$) and Bianchi II ($\alpha = 1$). The solutions 
to the effective equations derived from this Hamiltonian were already
studied in \cite{CM-bianchi2}, and we shall use them
to study the solutions of the effective Bianchi IX models.
\medskip

The two effective Hamiltonians of Bianchi IX (\ref{H-effec-BIX2},\ref{H-effec-BIX4})
can be seen as the generalization of the two effective theories for the closed FLRW ($k=1$),
that are obtained using the same quantization method for the curvature and the two 
lapses $N=V$ and $N=1$ \cite{CK,CK-2}. 
The effective Hamiltonian for the closed FLRW model with $N=V$ is given by \cite{closed},
\be
\label{k1-2}
\mc{C}_{\rm H_{k=1}}^{(1)}=\frac{3V^2}{8\pi G\gamma^2 \lambda^2}
\big[\sin^2\lambda\beta-2D\sin\lambda\beta+(1+\gamma^2)D^2\big]-\frac{p_\phi^2}{2}\approx 0 \, ,
\ee 
with $D=\lambda\vartheta/V^{-1/3}$.
The effective Hamiltonian with $N=1$ is given by \cite{CK-2},
\be 
\label{k1-4}
\mc{C}_{\rm H_{k=1}}^{(2)}=\frac{3A(V)V}{8\pi G\gamma^2 \lambda^2}
\big[\sin^2\lambda\beta-2D\sin\lambda\beta+(1+\gamma^2)D^2\big]-\frac{p_\phi^2}{2}\approx 0  \, .
\ee 
The quantum theory from which this effective Hamiltonian comes includes more 
negative powers of the volume, i.e., inverse triad corrections. A more
detailed discussion about these quantizations and the solutions to their
effective equations can be found in \cite{CK,CK-2,Asieh}.

The variables $(V,\beta)$ are related with $(c,p)$ by \cite{slqc}:
\be 
V=p^{3/2}\,,\quad\beta = \f{c}{ \sqrt{p} }\,.
\ee 
The Poisson brackets are
\be 
\{\beta,V\}=4\pi G \gamma\,\,,\quad  \{\phi,p_\phi\}=1\,.
\ee 
Finally, we want to consider the effective Hamiltonian for the flat FLRW model $k=0$,
since it is a limit of Bianchi IX spacetime and can be useful to check the correct numerical 
implementation. Furthermore, it can also be used to understand the behaviour of 
Bianchi IX near to the isotropic universes. The effective Hamiltonian for FLRW with $k=0$
is given by
\be
\mc{C}_{\rm H_{k=0}}=\frac{3}{8\pi G\gamma^2\lambda^2}\,V^2 \,\sin(\lambda\beta)^2 -\f{p_\phi^2}{2}\approx 0\, ,
\ee
with lapse $N=V$, and $V$ the physical volume of the fiducial cell.

Once one gets all the effective theories, the next step is to study their numerical
solutions and use them to classify and understand the Bianchi IX dynamics. 
This will be reported in the next section.

\section{Numerical Solutions}
\label{sec:4}

In this section we study the numerical solutions to the effective equations 
of the LQC Bianchi IX model. We consider two effective theories of Bianchi IX, one with
lapse $N=V$ \cite{bianchiIX}, and another with $N=1$ \cite{CK-3}. Given the difficulty to study
the solutions of these theories our strategy follows two paths. The first one is the discussion of 
the two effective theories from the analytical point of view.
Second, we study the space of solutions where the two theories describe
the same physics and justify the relevance of this region in solution space. Next, we consider several regimes that the solutions space may have, in order to exhibit different types of behaviour. We do not expect those cases to be exhaustive, but they give us important, physical, information. In particular, we
consider the Bianchi I limit, the non-shear limit and the isotropic limit, within this set of solutions.
\smallskip

One of the objectives of this part is to answer some questions regarding the effective solutions. 
First, we would like to know whether the 
singularity is resolved and whether the solutions reduce to the classical ones far from the bounce.
As we shall describe in detail, the answer is in the affirmative.
Second, we would like to verify that Bianchi IX reduces to the closed FLRW ($k=1$) when $p_1=p_2=p_3$ and $c_1=c_2=c_3$.
Third, we study the limit of flat FLRW ($k=0$) when $p_1=p_2=p_3$, $c_1=c_2=c_3$ and 
$x_1, x_2, x_3\rightarrow 0$.
Finally, we investigate whether the LQC Bianchi IX model has (LQC) Bianchi I as a limit when 
$x_1, x_2, x_3\rightarrow 0$.

In order to answer the second question we must discuss the differences
between the two effective theories from the analytical point of view.
It can be shown \cite{Asieh} that the Hamiltonian $\mc{C}_{\rm H_{BIX}}^{(1)}$ for Bianchi IX with 
lapse $N=V$, Eq. \eqref{H-effec-BIX2}, reduces to the effective closed FLRW with 
$N=V$, Eq. \eqref{k1-2}, when $p_1=p_2=p_3$ and $c_1=c_2=c_3$, in which case 
$\mathcal{C}_{\rm H_{BIX}}^{(1)}=\mc{C}_{\rm H_{k=1}}^{(1)}$. This means that,
when one quantizes both the isotropic and the anisotropic models, the solutions
to the anisotropic effective model must contain the solutions of the isotropic effective model. 
For the full quantum theory, one expects that the quantization of a less 
symmetrical model reduces to the quantization a more symmetrical one, as is explicitly the case for
the Bianchi I model \cite{bianchiI}. In the case of Bianchi IX such
reduction is still to be shown at the quantum level. 
On the other hand, the new quantization for Bianchi IX with $N$=1 and inverse corrections put forward in
\cite{CK,Asieh}, has a different behaviour. When we consider its effective Hamiltonian $\mc{C}_{\rm H_{BIX}}^{(2)}$, Eq.~\eqref{H-effec-BIX4}, and choose 
$p_1=p_2=p_3$, $c_1=c_2=c_3$, it does {\it not} reduce to the closed FLRW with 
$N=1$, Eq. \eqref{k1-4}, namely $\mathcal{C}_{\rm H_{BIX}}^{(2)}\ne \mc{C}_{\rm H_{k=1}}^{(2)}$.
Therefore, the dynamics that describe the two Hamiltonians for the isotropic case are
different. In this case the process of quantization and the symmetry reduction do not
commute. A more detailed study regarding the differences of these two theories at the quantum 
and effective levels can be found in \cite{Asieh,CK-3}.
\medskip

As the detailed study of the closed FLRW model with different quantizations
\cite{CK,CK-2} has shown, their effective theories, while providing  different dynamics for the bounce, 
yield the same results in the limit of large volume or large momentum of the field. 
The natural question is whether such feature extends to the anisotropic scenario.
With this question in mind is that we study the numerical solutions of Bianchi IX. The idea is to look for
solutions in the limit 
of large momentum of the field, i.e., where the volume at the bounce is expected to be much larger
than the Planck volume (as is the case for the $k$=0,1 models). 
It is important to recall that in both the closed FLRW and the Bianchi IX model
it is valid to talk about large volumes, since the spatial manifold is compact and has a finite physical volume.
This does not happen in neither of the Bianchi I, II and open FLRW models
where the quantity $V$ represents the physical volume of the fiducial cell.

\medskip

This section contains five parts. In the first one, we study the limit of large field momentum
and show that the two effective theories of Bianchi IX reproduce the same dynamics. 
In the second part, we show that the effective solutions resolve the big bang singularity, 
and reproduce the classical solutions far from the bounce. In the third one, we
explore the isotropic limits, for both flat and closed FLRW models. This allows us
to compare the numerical solutions with the known analytical and numerical results for
the isotropic models. In the fourth part, we consider the limit of small shear, which 
shows a very different behaviour than the isotropic limit. 
In the final part, we study the Bianchi I limit, explore  the 
transitions and discuss new results concerning to Bianchi IX and Bianchi VII${}_0$.
Throughout the units used to perform the simulations are:  
$c=1, \hbar=1, G=1, \gamma = 0.23753295796592$.
The numerical method used to integrate the equations is a Runge-Kutta 4 
method.\footnote{The program is available up on request.}
All the plots show simulations done with the cosmic time (lapse $N=1$).
When we choose $t=0$ as the time that the bounce occurs, 
we should clarify that this is the bounce in the geometric mean of the scale
factors, or equivalently when the expansion $\theta$ (or average Hubble rate)
vanishes.

\subsection{Large Field Momentum Limit}
\label{sec:big-vol}

\begin{figure}[tpb!]
\begin{center}
\begin{tabular}{c}
\includegraphics[width=0.5 \textwidth]{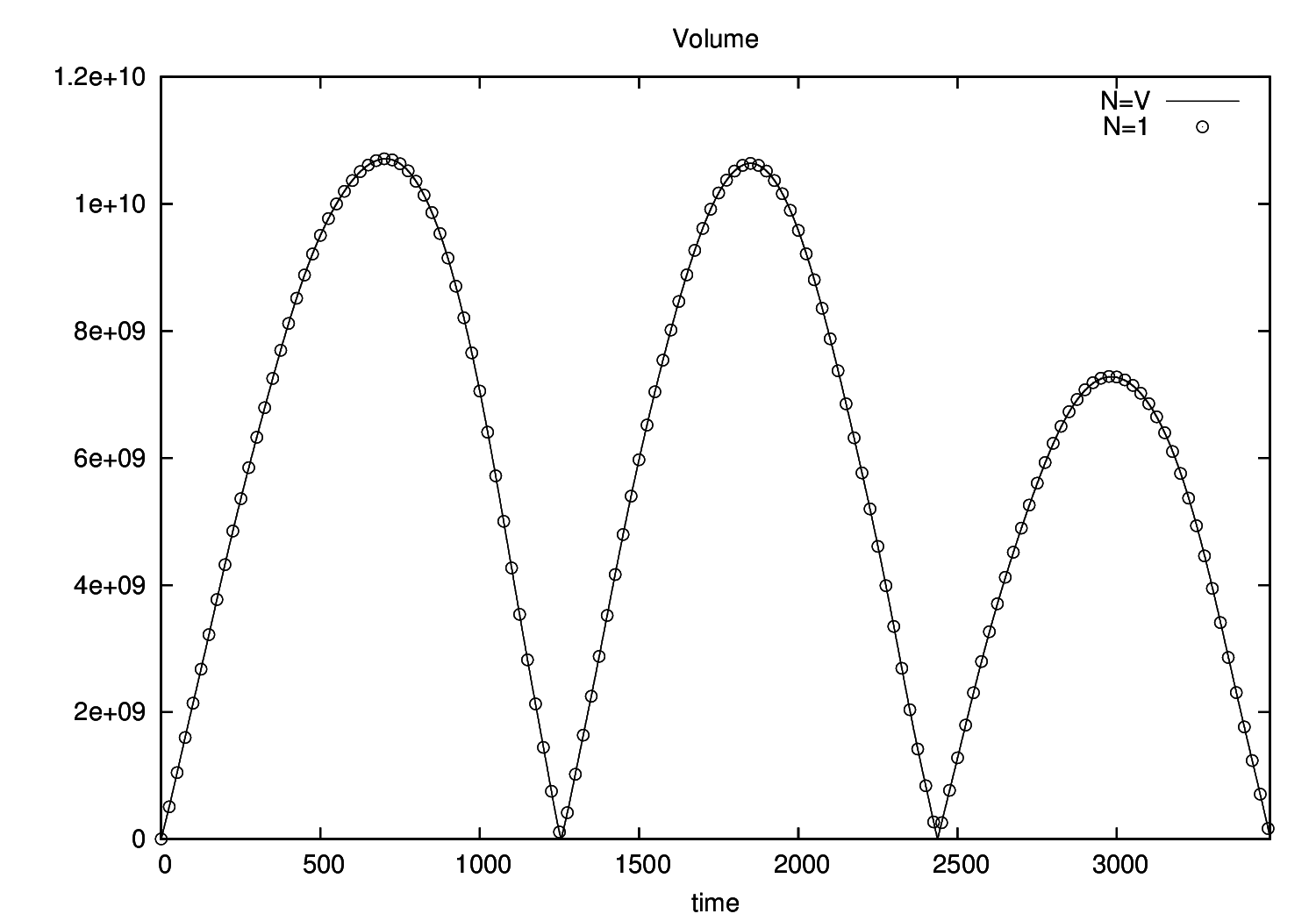}
\includegraphics[width=0.5 \textwidth]{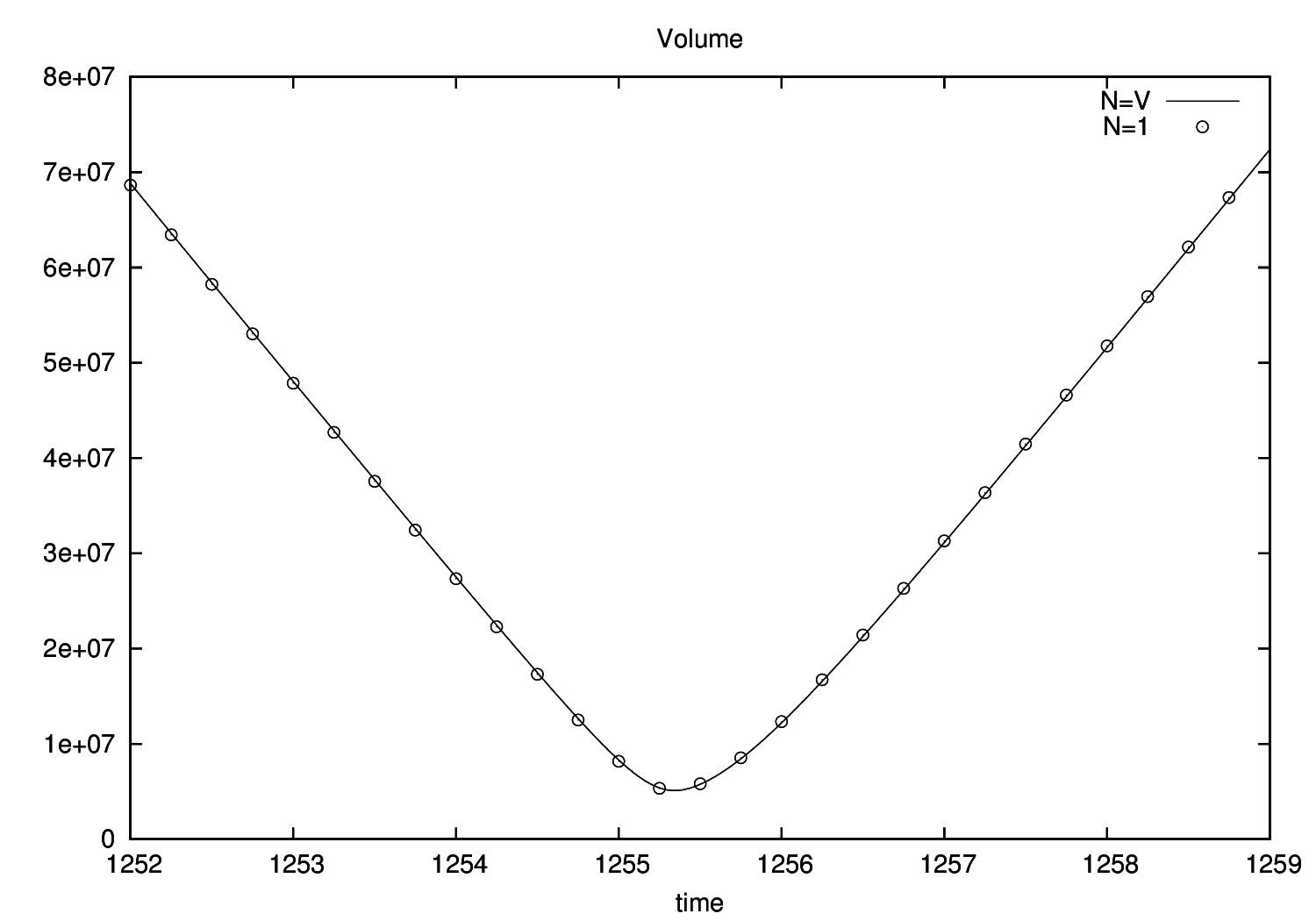}
\end{tabular}
\caption{Time evolution of the volume for the two  different effective theories, with lapses $N$=$V$ and $N$=1.
The initial conditions are: $\bar\mu_1 c_1 = \pi/3$, $\bar\mu_2 c_2= \pi/2$, $\bar\mu_3 c_3=\pi/4$, 
$p_1=13000$, $p_2=27000$ and $p_3=42000$. As it can be seen from the left figure, the universe undergoes 
a series of bounces and recollapses, and the dynamics of both theories is indistinguishable. In the right figure we zoom into one of the bounces where the dynamics of both theories coincides again.}
\label{fig:b9-Ed-A-vol}
\end{center}
\end{figure}

\begin{figure}[tpb!]
\begin{center}
\begin{tabular}{c}
\includegraphics[width=0.5 \textwidth]{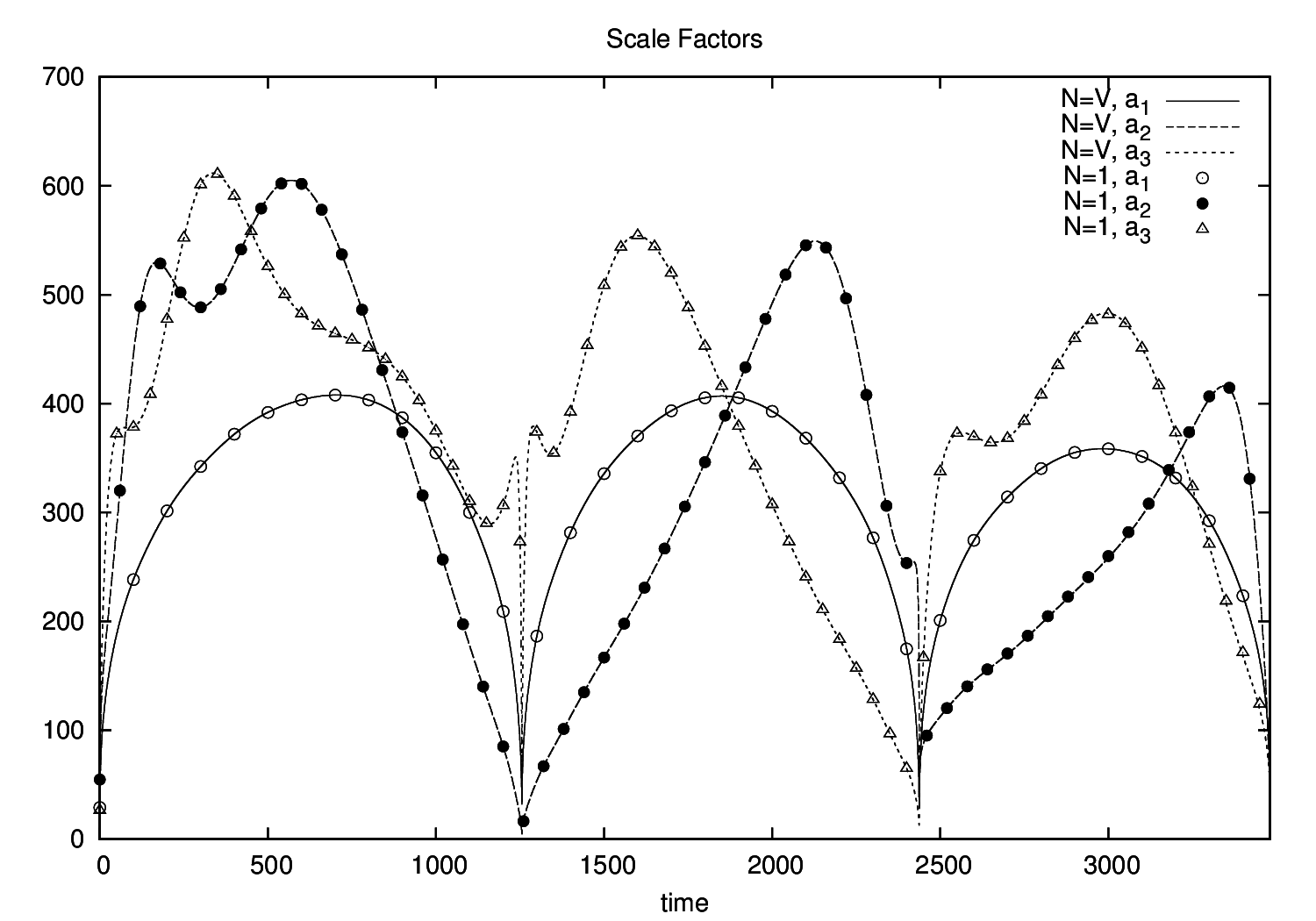}
\includegraphics[width=0.5 \textwidth]{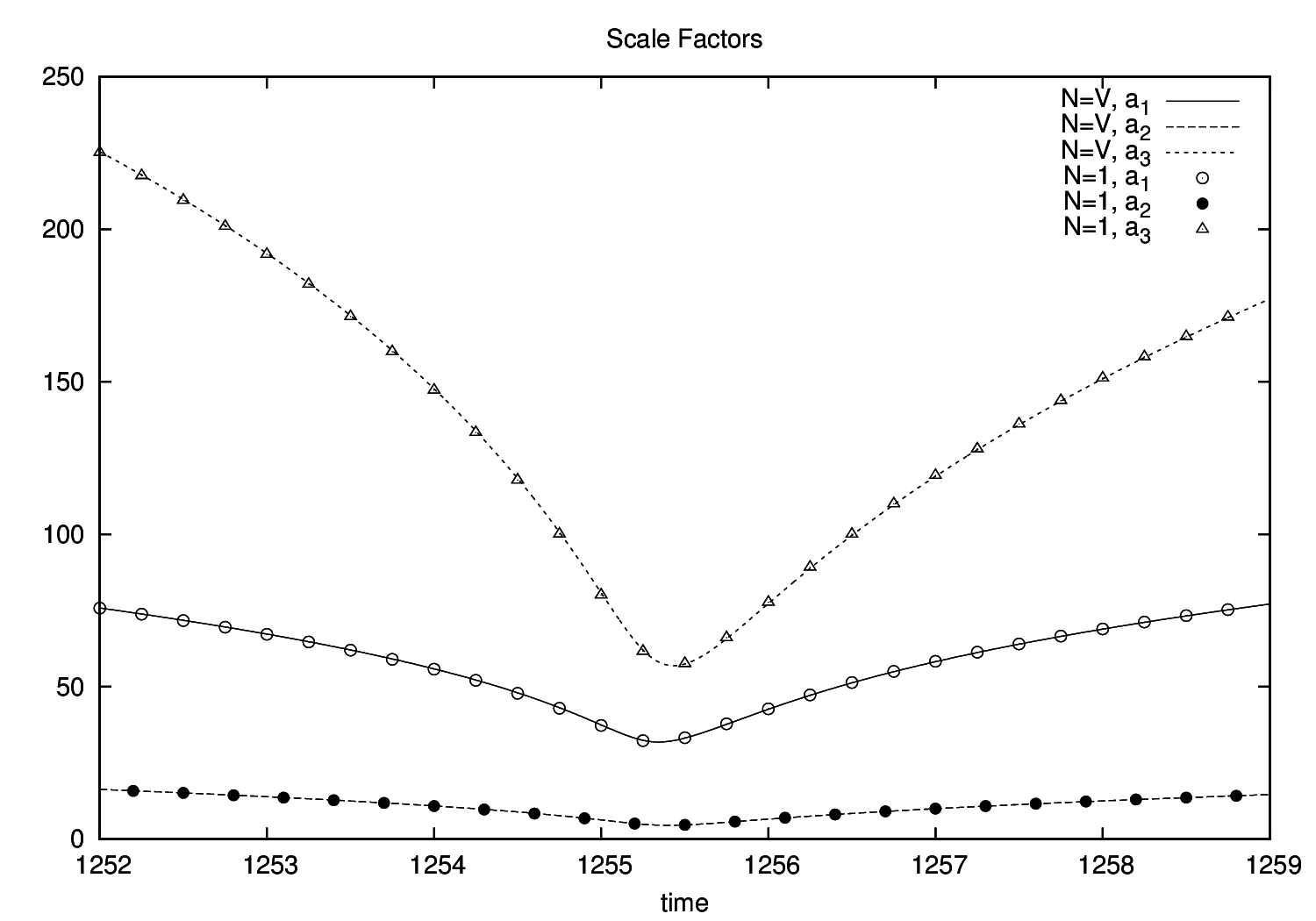}
\end{tabular}
\caption{Time evolution of the directional scale factors ($a_1,a_2,a_3$) for the two  different effective theories,
 with lapses $N$=$V$ and $N$=1. The initial conditions are: $\bar\mu_1 c_1 = \pi/3$, $\bar\mu_2 c_2= \pi/2$, $\bar\mu_3
  c_3=\pi/4$, $p_1=13000$, $p_2=27000$ and $p_3=42000$. Note that the dynamics of both theories coincides, even through the bounce (right).}
\label{fig:b9-Ed-A-scale-factors}
\end{center}
\end{figure}

\begin{figure}[tpb!]
\begin{center}
\begin{tabular}{c}
\includegraphics[width=0.5 \textwidth]{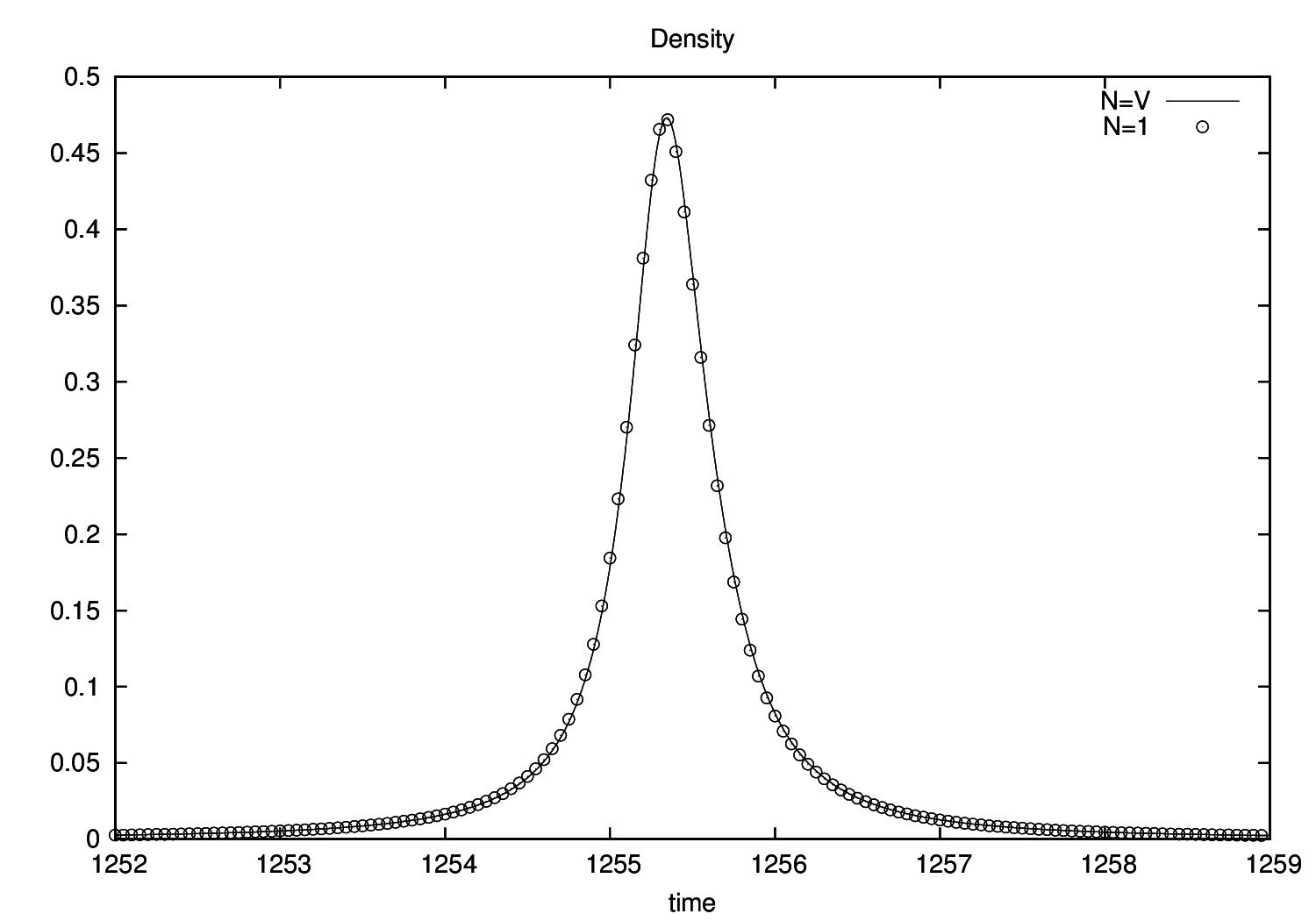}
\includegraphics[width=0.5 \textwidth]{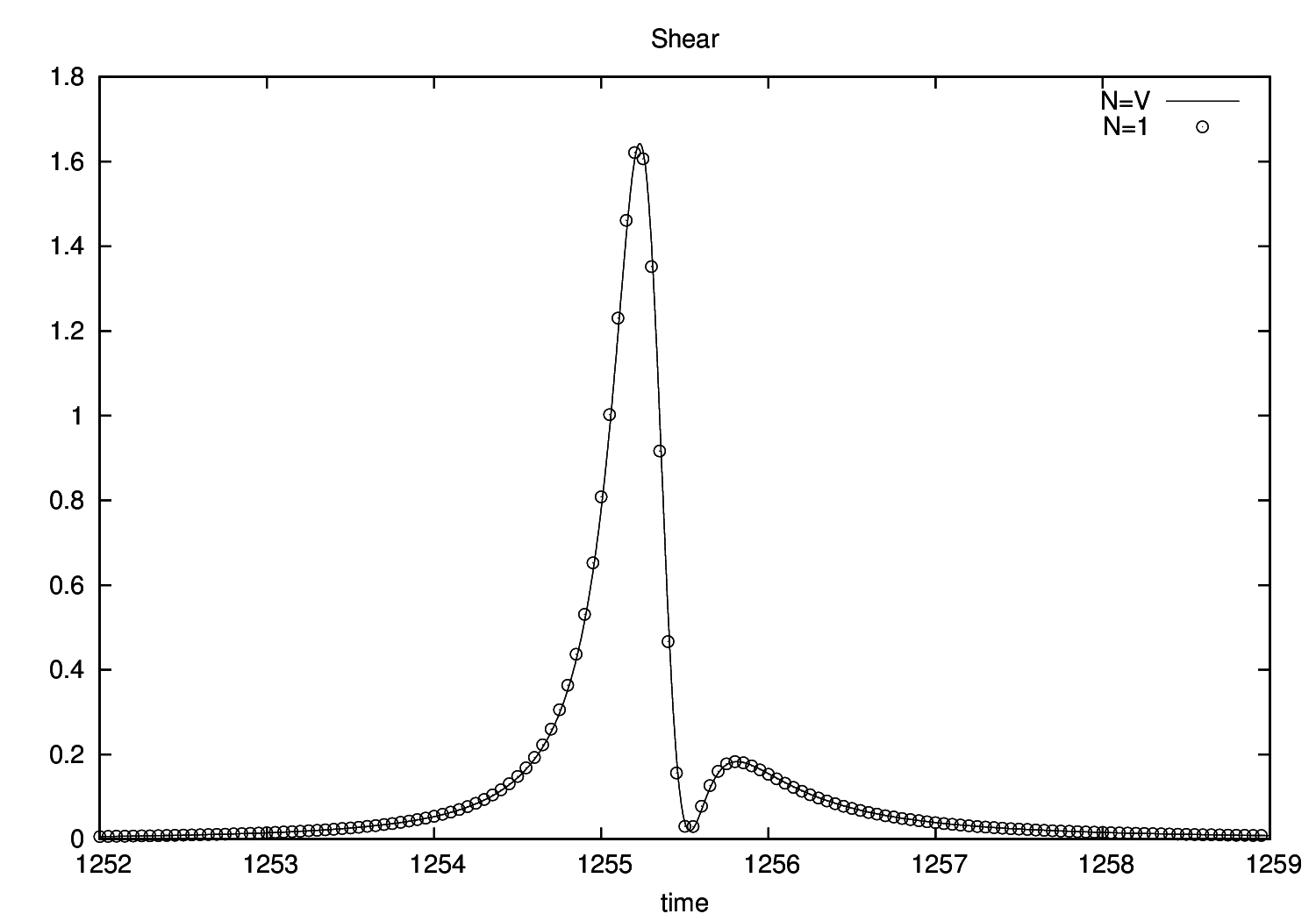}
\end{tabular}
\caption{Time evolution of the density and shear for the two different effective theories, with lapses $N$=$V$ and
$N$=1. The initial conditions are: $\bar\mu_1 c_1 = \pi/3$, $\bar\mu_2 c_2= \pi/2$, $\bar\mu_3 c_3=\pi/4$, 
$p_1=13000$, $p_2=27000$ and $p_3=42000$. Again, both theories can not be distinguished, even in the Planck regime.}
\label{fig:b9-Ed-A-dens-shear}
\end{center}
\end{figure}

First we are going to study the region in the space of solutions
where both effective equations give the same dynamics. 
This region correspond to the solutions with large field momentum. 
In this limit we can say 
that the two theories are equivalent, in the sense that, given an interval
of time $t_f-t_i<\infty$ and a tolerance $\epsilon>0$, it is always 
possible to find initial conditions such that the solutions of both theories 
evolve within the tolerance $\epsilon>0$ for the interval of time $t_f-t_i<\infty$.
It is important to clarify that all the assertions about the effective solutions
apply to the numerical solutions that we study, and not for all the space of
solutions. Nevertheless, we expect that the qualitative results found in our numerical explorations
can be extended to other, generic within the respective class, initial data.

In Figs.~\ref{fig:b9-Ed-A-vol} ,\ref{fig:b9-Ed-A-scale-factors}, \ref{fig:b9-Ed-A-dens-shear}
the comparison between the solutions of both effective theories can be appreciated: the three scale factors
have the same behavior in both theories. The bigger the
volume at bounce is (or the momentum of the field), the closer the two effective
solutions are. The initial conditions are: 
$\bar\mu_1 c_1 = \pi/3$, $\bar\mu_2 c_2= \pi/2$, $\bar\mu_3 c_3=\pi/4$, 
$p_1=13000$, $p_2=27000$ and $p_3=42000$.
The momentum of the field is $p_\phi=3.17\times 10^{6}$, that comes from the 
Hamiltonian constraint. 

From now on, the study of the effective solutions will be at the regime of
large field momentum. Since we have shown that in that limit both effective theories are 
indistinguishable, all the conclusions that we reach will apply for both effective 
theories of the Bianchi IX model. In particular,   the big bang is replaced by a bounce in all cases. 
This will be the subject of study in the next section, together with the comparison with the classical 
solutions.

\subsection{Classical Limit}
\label{sec:lim-clas}

\begin{figure}[tpb!]
\begin{center}
\begin{tabular}{c}
\includegraphics[width=0.5 \textwidth]{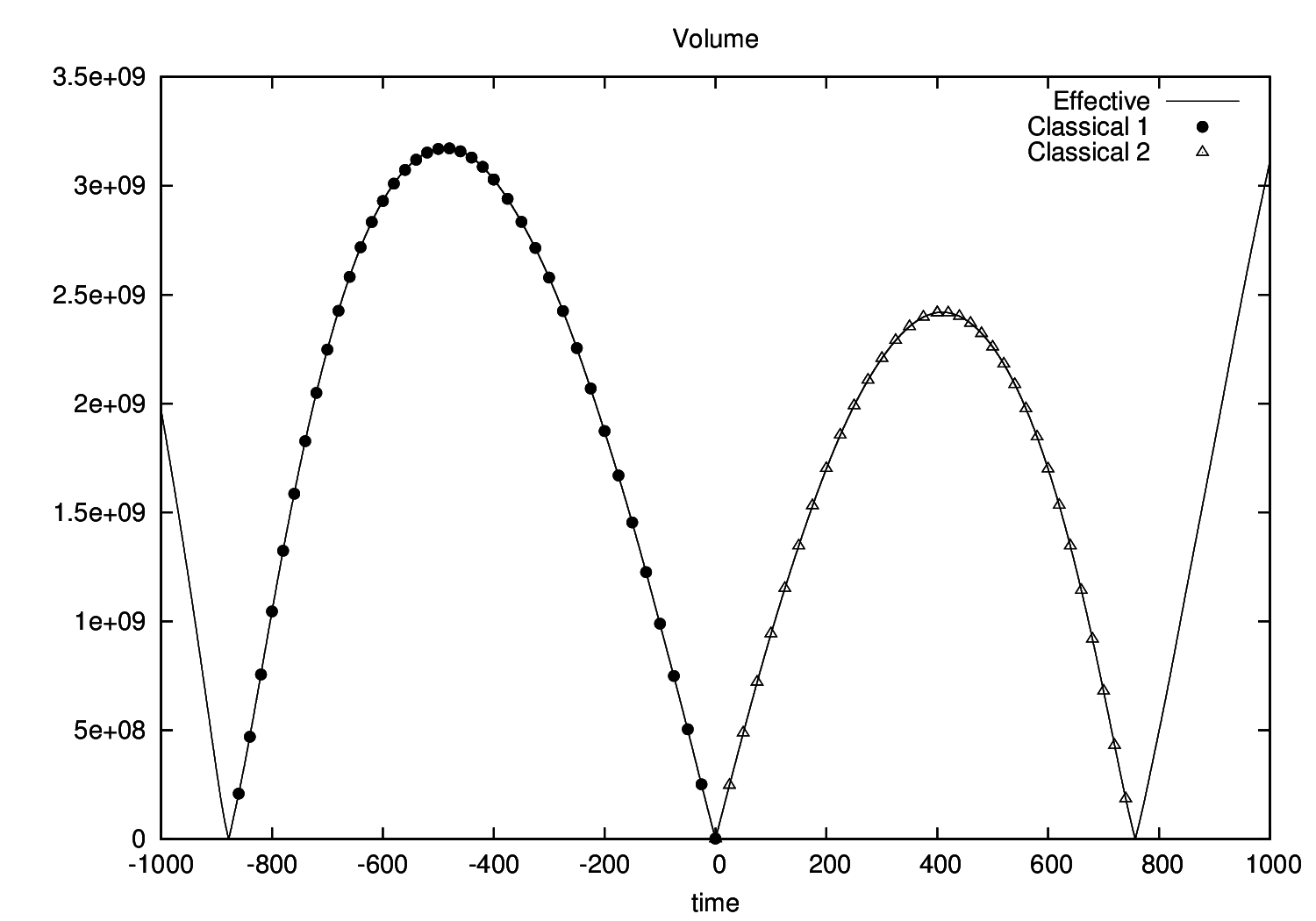}
\includegraphics[width=0.5 \textwidth]{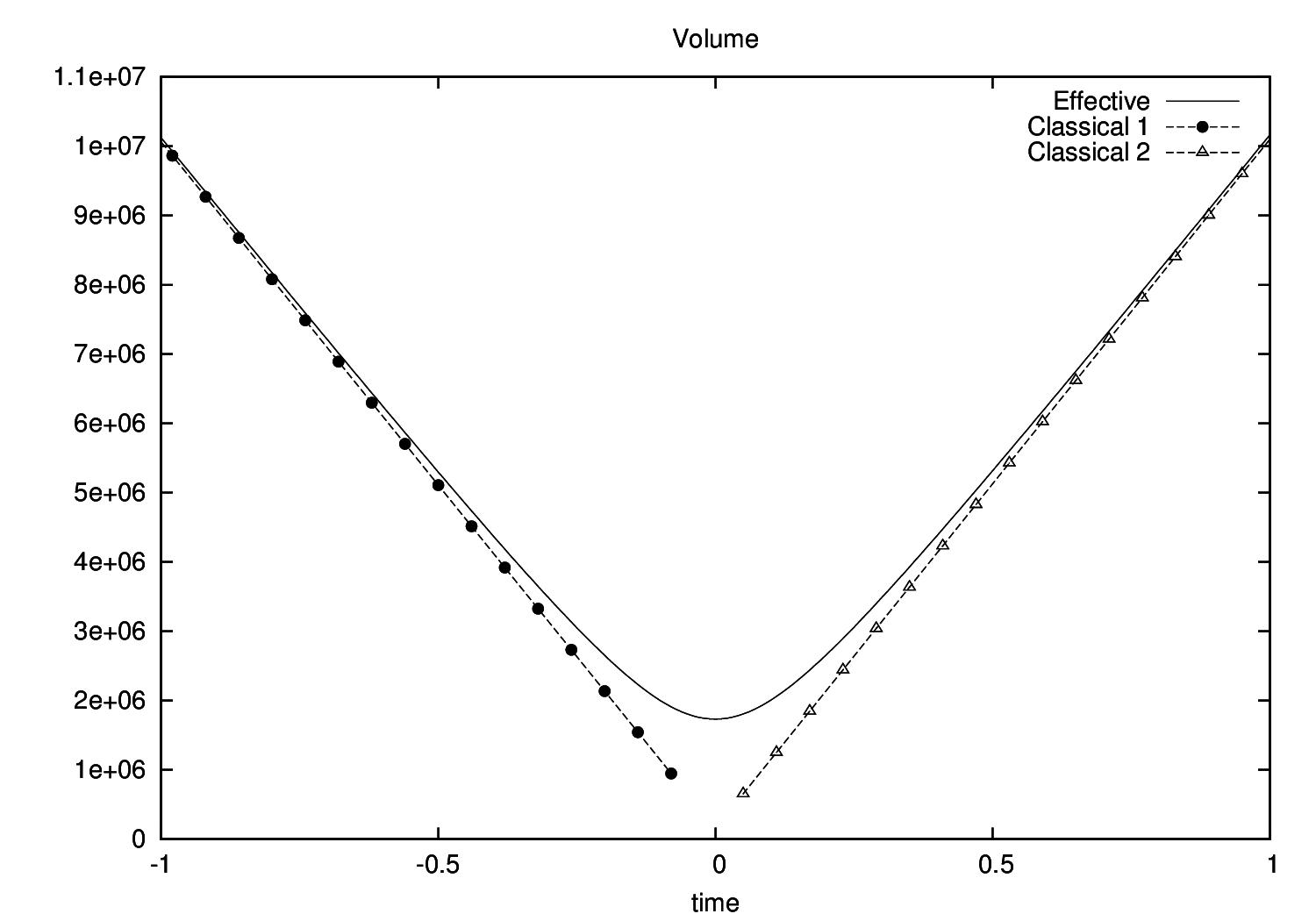}
\end{tabular}
\caption{Time evolution of the volume. Here we compare effective solution with the classical solutions they approach,
before and after the bounce (that occurs at $t=0$). It can be appreciated that the classical and effective solutions
have the same behaviour in the classical region. Near to the bounce the classical solutions go to 
zero volume (big bang singularity) and the effective solution bounces. The initial conditions are: 
$\bar\mu_1 c_1= 3\pi/8$, $\bar\mu_2 c_2= \pi/2$, $\bar\mu_3 c_3=5\pi/8$, $p_1=15000$, $p_2=10000$, $p_3=20000$.
The initial conditions for the classical equations are taken from the effective evolution at the maximal volume, and evolved backwards.}
\label{fig:class-b9-volume}
\end{center}
\end{figure}

\begin{figure}[tpb!]
\begin{center}
\includegraphics[width=11cm]{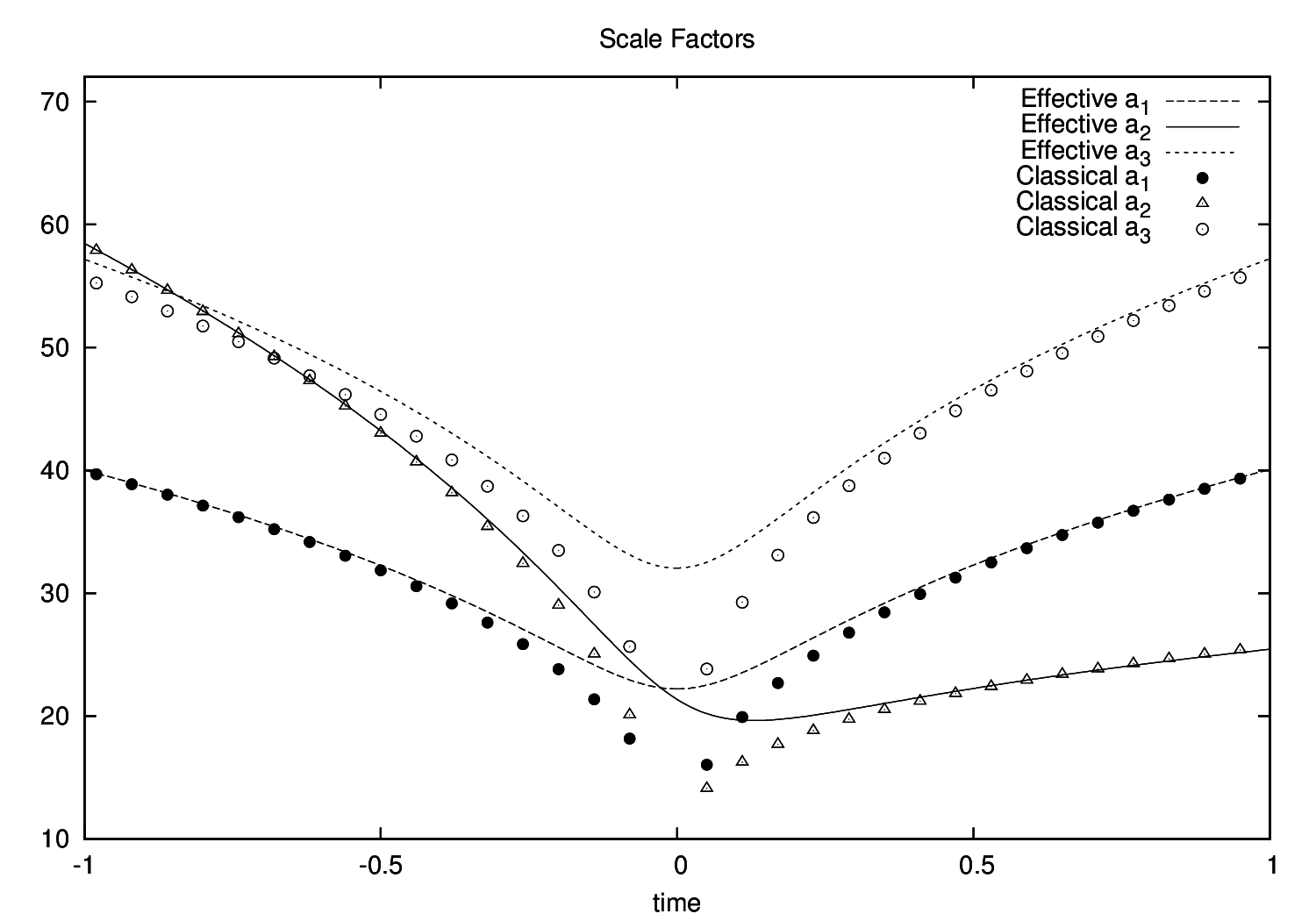}
\caption{Time evolution of the scale factors ($a_1,a_2,a_3$). As in the previous figure, we compare the effective
and classical solutions. 
It can be appreciated that each direction bounces (at a different time)
in the effective solution, while all the classical solutions go to zero at some point, either before and after the
bounce (that occurs at $t=0$). The initial conditions are: 
$\bar\mu_1 c_1= 3\pi/8$, $\bar\mu_2 c_2= \pi/2$, $\bar\mu_3 c_3=5\pi/8$, $p_1=15000$, $p_2=10000$, $p_3=20000$.
The initial conditions for the classical equations are taken from the effective evolution at the maximal volume.}
\label{fig:class-b9-scale-factors}
\end{center}
\end{figure}

\begin{figure}[tpb!]
\begin{center}
\begin{tabular}{c}
\includegraphics[width=0.5 \textwidth]{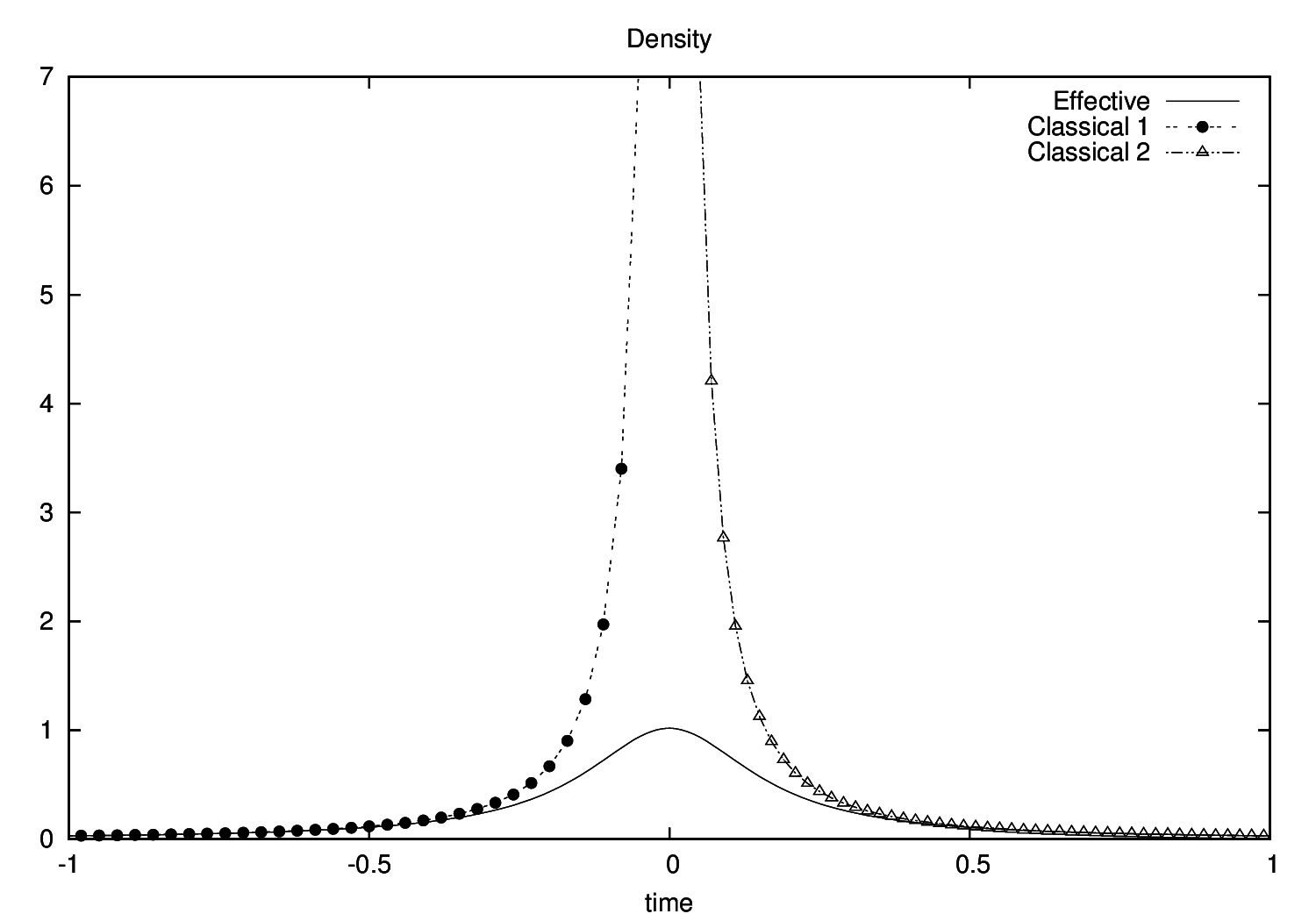}
\includegraphics[width=0.5 \textwidth]{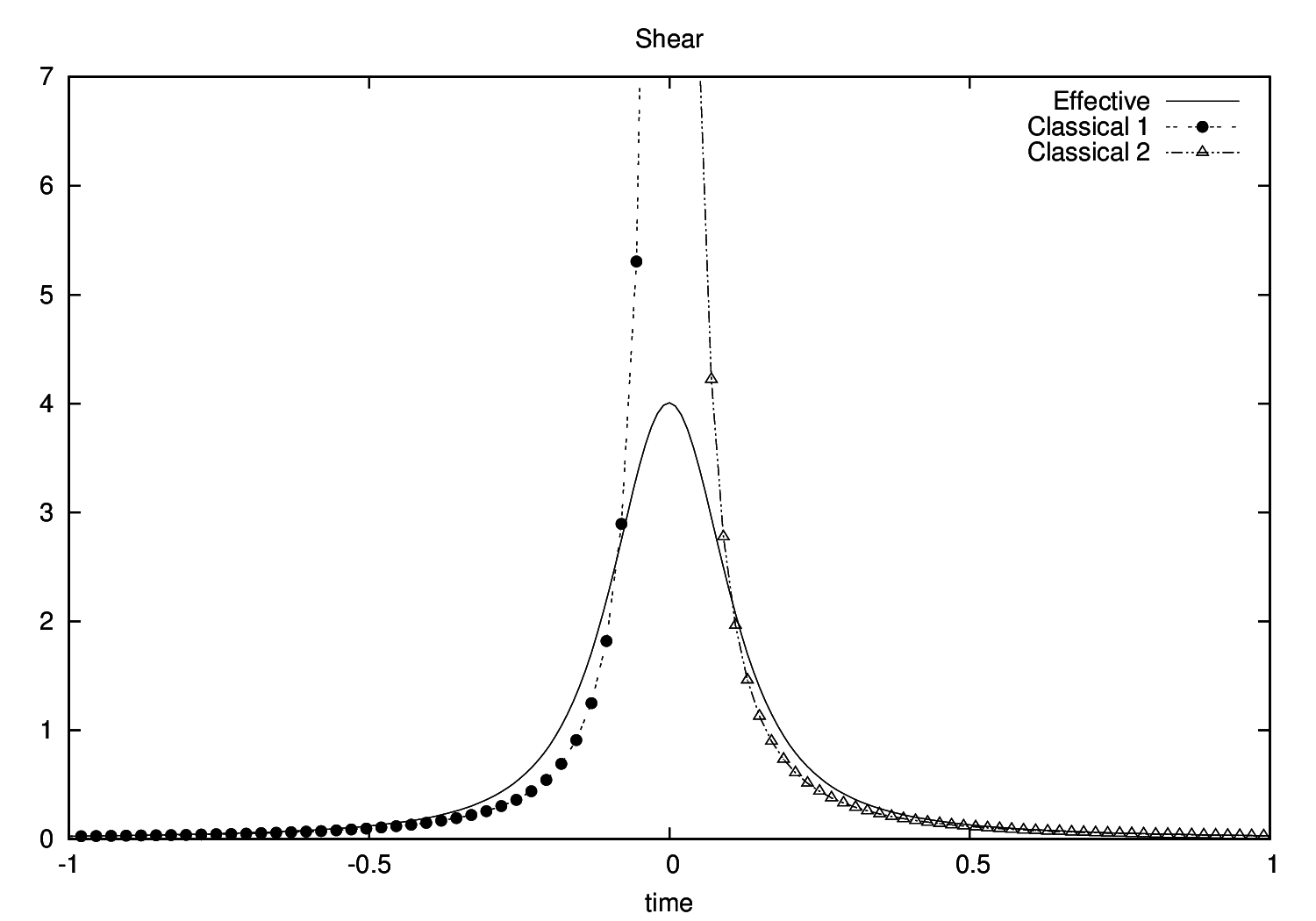}
\end{tabular}
\caption{Time evolution of the density and shear. The effective and classical solutions are compared.
It can be appreciated that, for the classical solutions the density and shear diverge, either as the
solution approaches a big crunch singularity if it is contracting, or possesses a big bang singularity if it is expanding. Both scalar quantities remain finite in the effective solution.
The initial conditions for the effective equations are: 
$\bar\mu_1 c_1= 3\pi/8$, $\bar\mu_2 c_2= \pi/2$, $\bar\mu_3 c_3=5\pi/8$, $p_1=15000$, $p_2=10000$, $p_3=20000$.
The initial conditions for the classical equations are taken from the effective evolution at the maximal volume.}
\label{fig:class-b9-dens-shear}
\end{center}
\end{figure}

We want to verify that the classical big bang
singularity is resolved by the effective dynamics and check that the classical dynamics is recovered 
in some interval of time. This add self-consistency to the effective theory, in the sense
that it reproduces the classical dynamics far from the bounce (consistent with general relativity), 
{\it and}  it solves the big bang singularity incorporating then what we expect to be the dynamics
of the semiclassical states of the quantum theory. \medskip

We choose the initial conditions 
near to the bounce for the effective equations and evolve them back and forward in 
time. The initial conditions are: 
$\bar\mu_1 c_1= 3\pi/8$, $\bar\mu_2 c_2= \pi/2$, $\bar\mu_3 c_3=5\pi/8$, $p_1=15000$, $p_2=10000$, $p_3=20000$.
In order to do an easy comparison with the classical solutions, we take the values
of the variables $(c_i,p_i)$ at the maximal volume, that come from the effective evolution, and
later on we introduce them as initial data for the classical equations, and then we evolve
back and forward in time.\medskip

The effective and classical solutions are compared in 
Fig.~\ref{fig:class-b9-volume}, \ref{fig:class-b9-scale-factors}, \ref{fig:class-b9-dens-shear}. 
Here we show the time evolution of the total volume and the scale factors. Note that the effective
and classical solutions coincide up to a point close to the bounce, then the classical
solutions have a singularity, while the effective solutions bounce and connect
with the classical solutions before and after the bounce. The recollapse is due to
the positive spatial curvature and that holds for any matter content that satisfies 
the dominant energy condition. 
The bounce is due to the loop quantum cosmology effects. Therefore the universe
continue bouncing and recollapsing forever. \medskip

Once we verify that the effective solutions reproduce the classical solutions
and solve the big bang singularity, the next step is to study the space of 
effective solutions. Since the dynamics of Bianchi IX is non trivial and
there are no analytical solutions for the classical dynamics, one is not
expected to get a complete and fully detailed study of the effective solutions. 
Still, it is possible to perform a qualitative study and get a better control 
on the space of solutions. For this reason we study different limits of the
solutions. One interesting limit is when the solutions are isotropic, which is the case 
we shall explore next.

\subsection{Isotropic Limit}
\label{sec:iso-lim}

\begin{figure}[tpb!]
\begin{center}
\includegraphics[width=11cm]{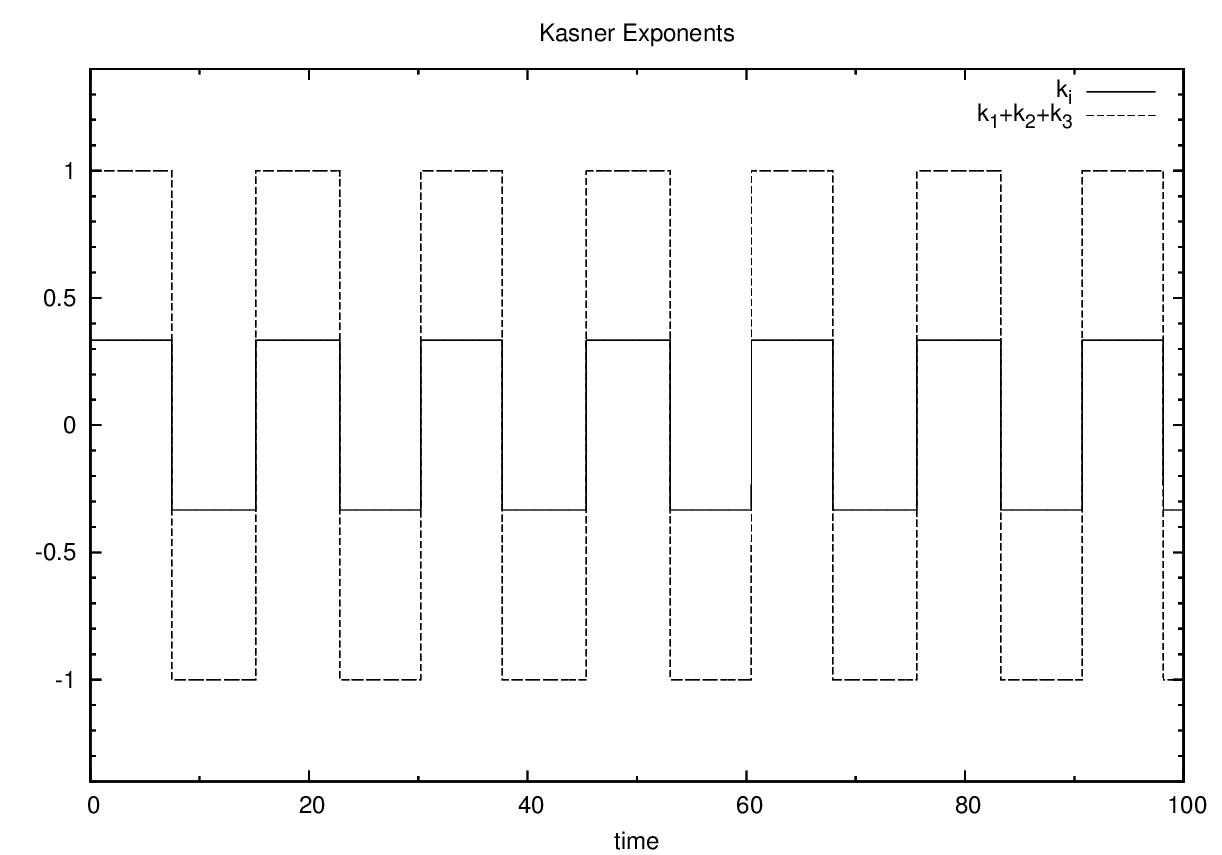}
\caption{Evolution of the Kasner exponents ($k_1,k_2,k_3$).
The initial conditions are: $\bar\mu_ic_i=\pi/2$, $p_i=100$, with $i=1,2,3$}
\label{fig:b9-iso-kasner}
\end{center}
\end{figure}

\begin{figure}[tpb!]
\begin{center}
\begin{tabular}{cc}
\includegraphics[width=0.5 \textwidth]{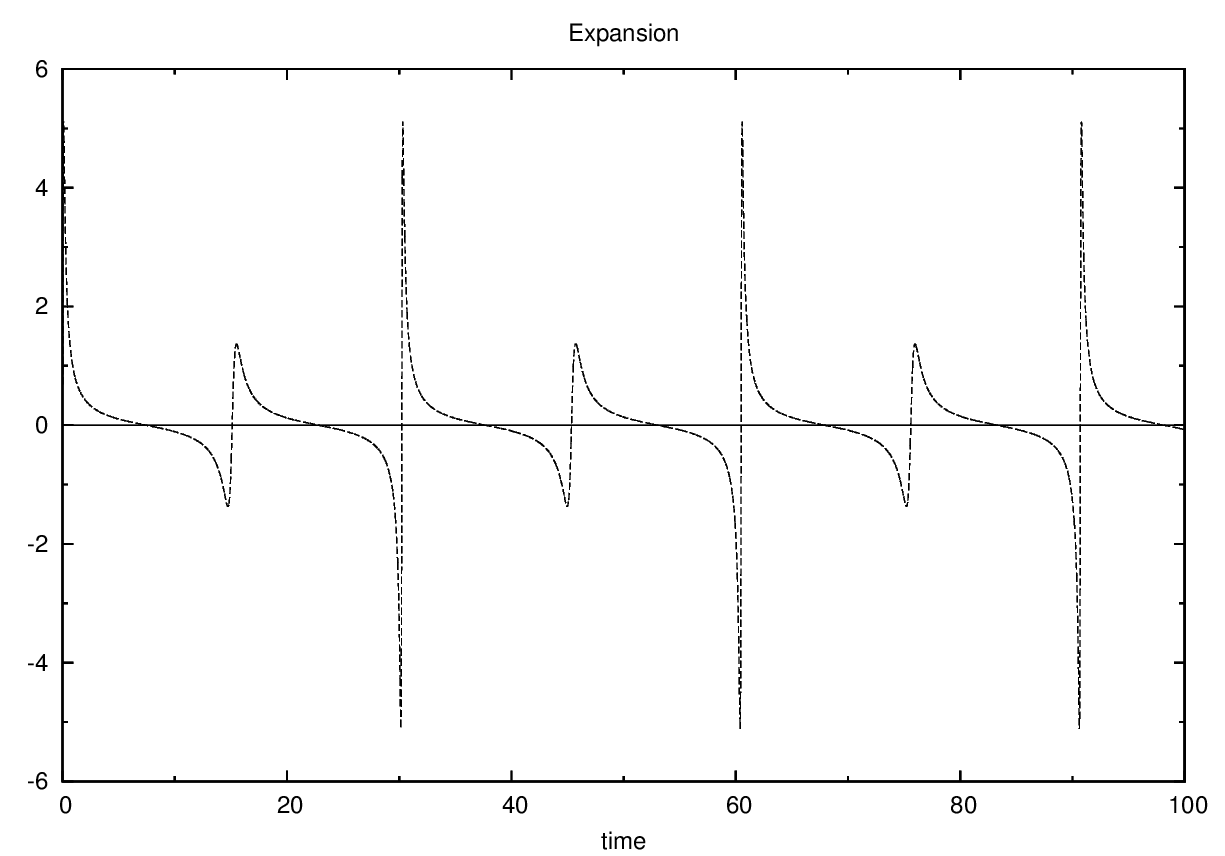} &
\includegraphics[width=0.5 \textwidth]{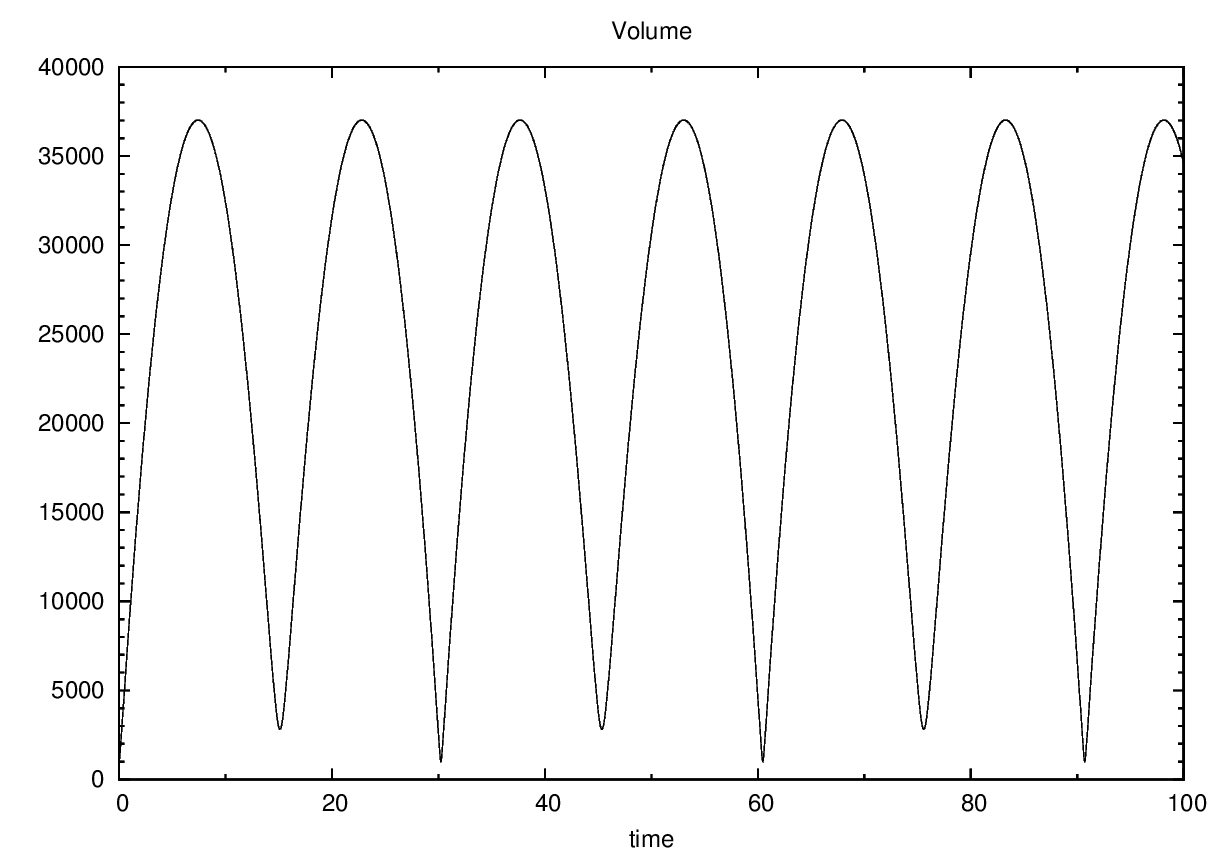}
\end{tabular}
\caption{Evolution of expansion and volume. 
The initial conditions are: $\bar\mu_ic_i=\pi/2$, $p_i=100$, with $i=1,2,3$.}
\label{fig:b9-iso-vol}
\end{center}
\end{figure}

\begin{figure}[tpb!]
\begin{center}
\includegraphics[width=11cm]{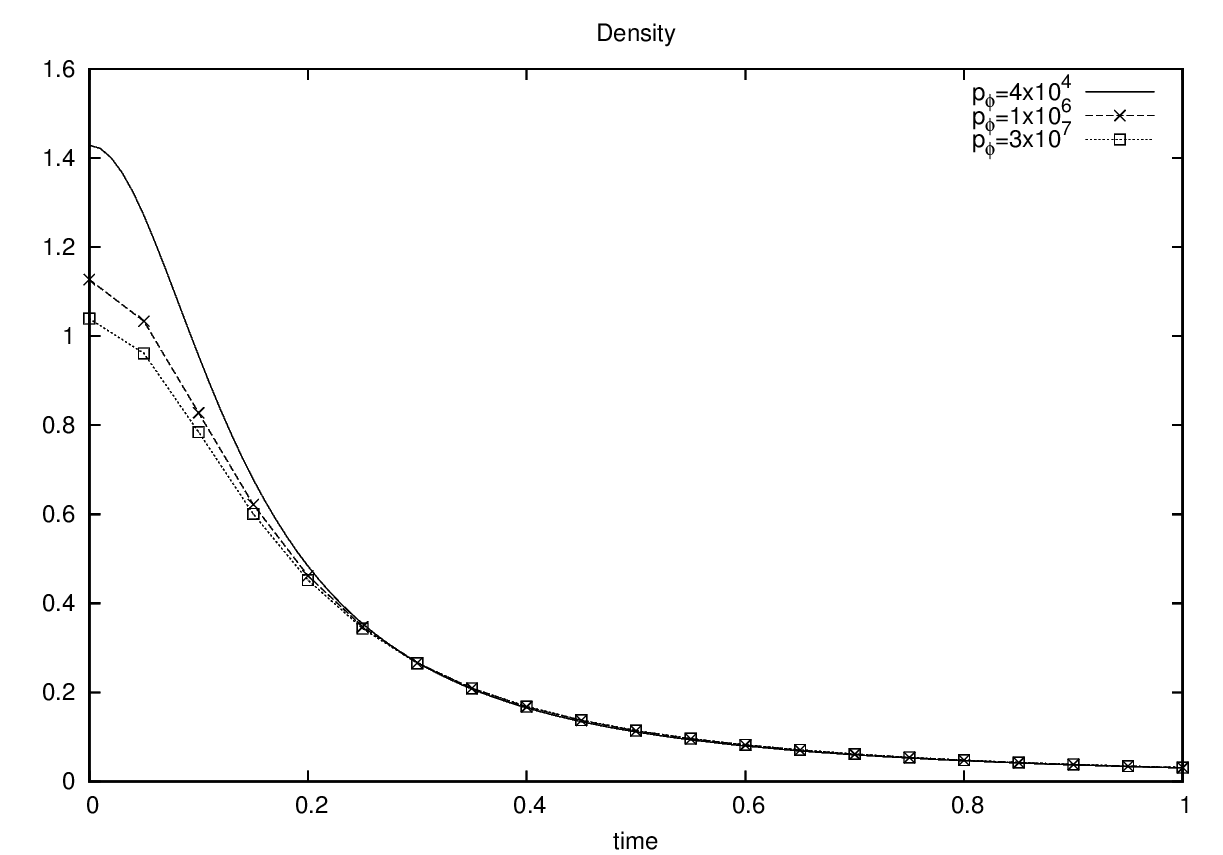}
\caption{Time evolution of the density ($\rho/\rho_{\rm crit}$) for solutions with different values for 
the momentum of the field $p_\phi$.
The bounce is at $t$=0. 
The initial conditions for the three solutions are: $\bar\mu_ic_i=\pi/2$, $p_i=1000$, $p_i=10000$, $p_i=100000$, with $i=1,2,3$. Note that as the momentum $p_\phi$ increases, the density at the bounce decreases and approaches the value of the $k$=0 FLRW model}
\label{fig:b9-iso-dens}
\end{center}
\end{figure}

Within the Bianchi IX solutions, the ones that are close to the isotropic solutions 
are very important, since it is expected that any `realistic' model of the universe must contain
these solutions. After all, our universe is almost homogeneous
and isotropic at large scales. Since the numerical solutions for the full quantum isotropic
universes are known \cite{aps0,aps,aps2,closed,singhnew}, 
they can be compared against the effective Bianchi IX solutions in the isotropic limit to test its
validity.
\medskip

In figure \ref{fig:b9-iso-kasner} we show the evolution of the Kasner
exponents $k_1,k_2,k_3$ and their sum $k_1 +k_2 +k_3$.  It can be seen from the figure
that all the exponents have the same value ($k_i=1/3$) and the total sum is one.
This proves that the expansion rate in each direction is the same.
The sign of the Kasner exponents indicates that its corresponding direction is expanding 
($k_i>0$) or contracting ($k_i<0$), while the sign of the sum $\pm 1$ tells us
the sign of the total expansion $\theta$. The bounce or the recollapse is when $\theta=0$,
if $\theta>0$ the universe is expanding and for $\theta<0$ it is contracting. 
This dynamics can be seen in figure \ref{fig:b9-iso-vol},
where it is shown the evolution of the volume and the expansion $\theta$. 
Note that there are two types of bounces. The volume at the bounce differs for
each type of bounce and it has two different values. This feature was first studied, for the
$k$=1 FLRW model in \cite{CK},
and correspond to the dynamics of the effective theory associated with a connection based quantization
(as opposed to the curvature based quantization of \cite{closed}). This 
allows us to verify the results for the isotropic $k=1$ case, and be confident
of our numerical results. The initial conditions are: $\bar\mu_ic_i=\pi/2$, $p_i=100$, with $i=1,2,3$.
\medskip

In figure \ref{fig:b9-iso-dens} we plot the density ($\rho/\rho_{\rm crit}$) as a function of cosmic time.
It shows how the density changes as the momentum $p_\phi$ of the field changes. One interesting thing to note
is that  
the density at the bounce ($t=0$) approaches the critical density ($\rho/\rho_{\rm crit}\approx 1$)
as the momentum of the field becomes larger, while the density of the bounce seems to increase as the
momentum decreases. Recall that we are in the isotropic limit, so the change in density at the bounce 
can be interpreted as a transition from the $k$=1 to the $k$=0 cases. 
Recall that the $k$=1 FLRW model possesses an absolute
bound on density only when inverse triad corrections are introduced, but the density at the bounce is 
always larger than in the $k$=0 case where the density is always bounded by  
$\rho_{\rm crit}$ \cite{CK,CK-2}.
Additionally, the time between each bounce is longer as the momentum of 
the field increases, since the universe reaches larger values of the volume and, therefore, it
takes more time to recollapse. Thus, as one performs simulations with larger values of the momentum of the field one approaches, as
a limiting case, the isotropic flat FLRW universe. This universe is not embedded within the 
Bianchi IX model but is rather a limit for the solutions, which can be approached as much as one wants 
in a finite interval of time. The initial conditions for each of the solutions 
in Fig. \ref{fig:b9-iso-dens} are:
$\bar\mu_i c_i=\pi/2$, $p_i=1000$, with $p_\phi\approx 4\times 10^4$; 
$\bar\mu_i c_i=\pi/2$, $p_i=10000$, with $p_\phi\approx 1\times 10^6$ and
$\bar\mu_i c_i=\pi/2$, $p_i=100000$, with $p_\phi\approx 3\times 10^7$.

Let us end this part by clarifying the situation at hand. 
Strictly speaking, we are not studying the isotropic limit corresponding to 
FLRW with $k=1$, given that we choose exactly isotropic initial conditions, while in
the $k=0$ case, it is truly a limiting case since the FLRW model is not contained
within the Bianchi IX model. Once
we have cleared up what we mean  by the isotropic limit, we proceed in the next section
to study the solutions
near to the isotropic limit, which we call the non-shear limit, since the 
shear has a small contribution throughout the dynamical evolution.

\subsection{Non-Shear Limit}
\label{sec:lim-nonshear}

\begin{figure}[tpb!]
\begin{center}
\begin{tabular}{cc}
\includegraphics[width=0.5 \textwidth]{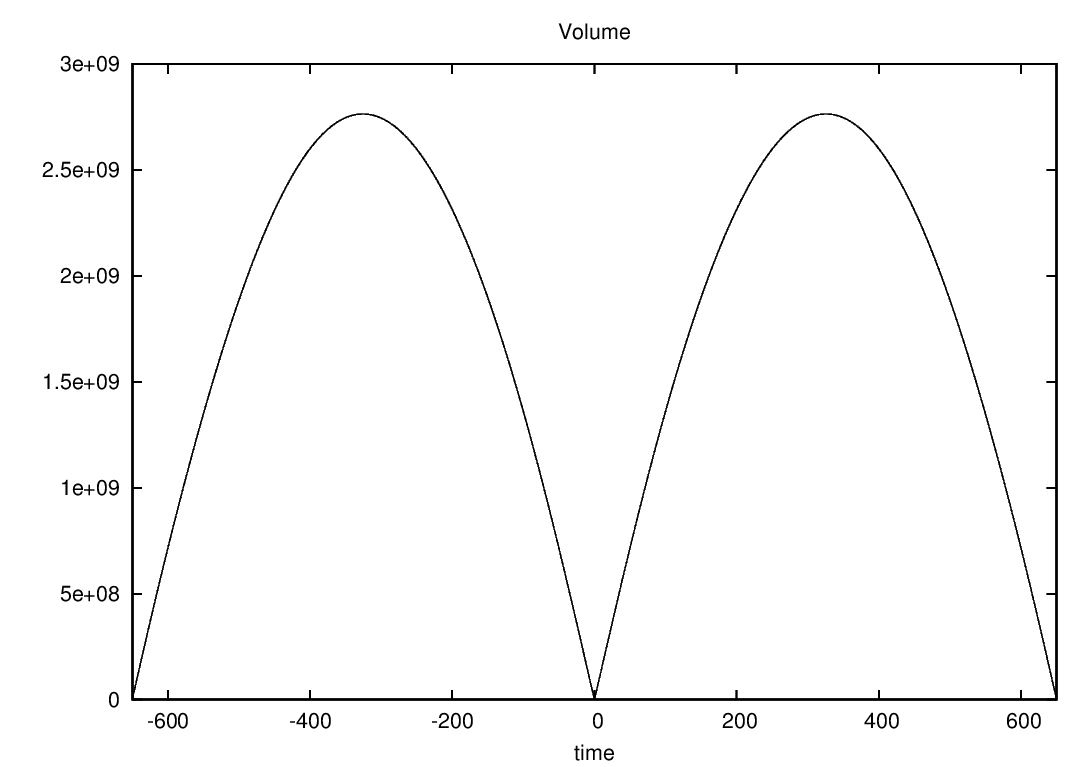} &
\includegraphics[width=0.5 \textwidth]{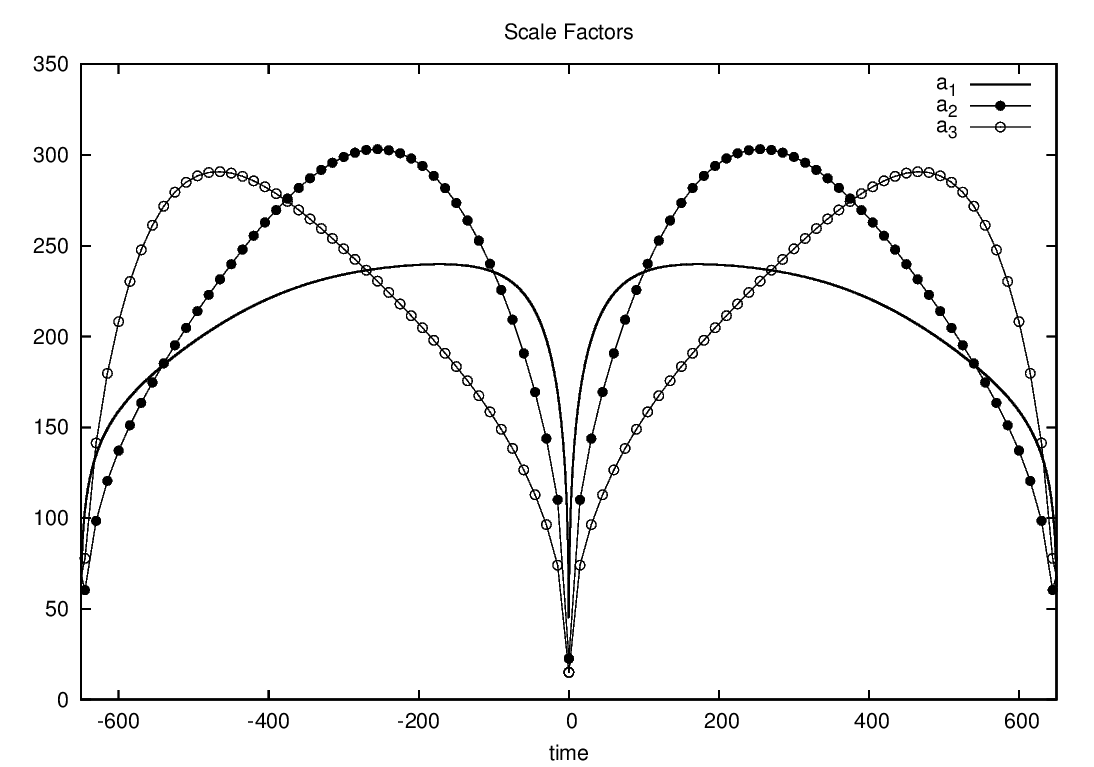}
\end{tabular}
\caption{Evolution of volume and scale factors. The initial conditions are: 
$\bar\mu_1 c_1=\bar\mu_2 c_2=\bar\mu_3 c_3=\pi/2$, $p_1=10000$, $p_2=20000$, $p_3=30000$.
}
\label{fig:nonshear-vol}
\end{center}
\end{figure}

\begin{figure}[tpb!]
\begin{center}
\begin{tabular}{cc}
\includegraphics[width=0.5 \textwidth]{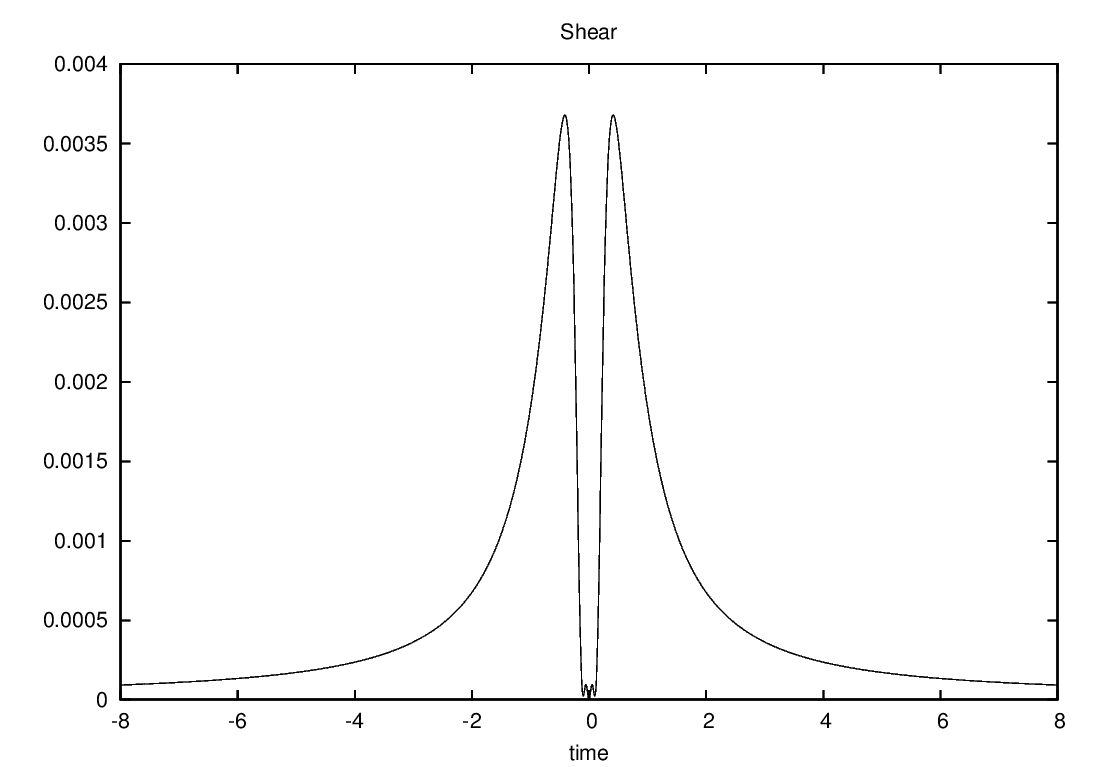} &
\includegraphics[width=0.5 \textwidth]{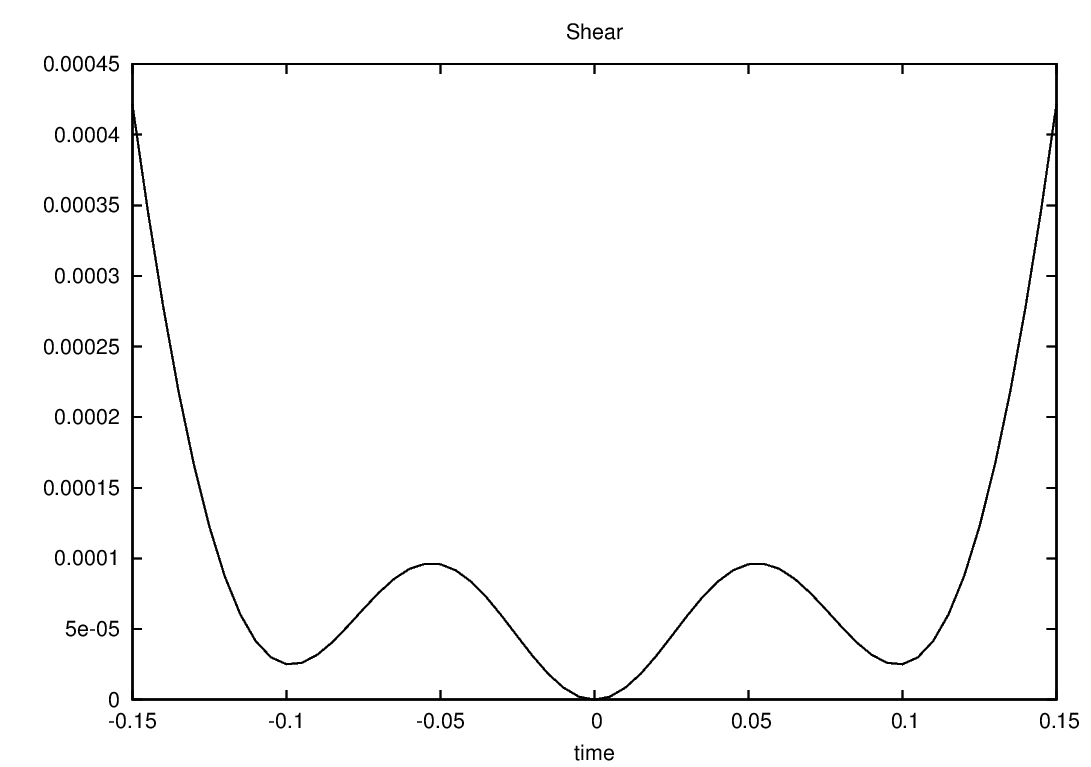}
\end{tabular}
\caption{During evolution the shear remains small and vanishes at the bounce due to the initial conditions: 
$\bar\mu_1 c_1=\bar\mu_2 c_2=\bar\mu_3 c_3=\pi/2$, $p_1=10000$, $p_2=20000$, $p_3=30000$.
}
\label{fig:nonshear-shear}
\end{center}
\end{figure}

In the previous section we clarified that the isotropic limit is really made of  
isotropic solutions between the Bianchi IX, 
given that we chose exactly isotropic initial conditions. Now, we want 
to study the evolution of full Bianchi IX near to the isotropic solutions,
namely, we want a small contribution from the shear. In order to do that, we
choose initial data that contain exactly zero shear at the bounce, and 
then study how the shear changes during time evolution.
While it is not a generic solution, it is an interesting limiting case.
Furthermore, it can also be used to test the accuracy of the numerical
code by checking whether the dynamics is symmetric across the bounce.
The dynamics is expected to be time-symmetric given that all velocities ($\dot{p}_i$) vanish 
at the bounce point, and therefore there is no difference in time evolution forwards or 
backwards. We choose the initial conditions 
at the bounce for the effective equations and evolve them back and forward in 
time, the initial conditions are: 
$\bar\mu_1 c_1=\bar\mu_2 c_2=\bar\mu_3 c_3=\pi/2$, $p_1=10000$, $p_2=20000$, $p_3=30000$.

Figure \ref{fig:nonshear-vol} shows the evolution of the volume 
and the scale factors. It is clear from the figure that the evolution is symmetric across the bounce.
This fact can be seen to validate the 
numerical code, but more intriguing, this solution is very different from the isotropic 
solution shown in the previous section. A new feature appears in the
figure \ref{fig:nonshear-shear} where it is shown the evolution of the shear, 
which vanishes at the bounce point, due to the initial conditions,
and is different from zero during the evolution. The shear has
a small value and the maxima are less than $4\times 10^{-3}$ 
(recall that the units are: $c=1, \hbar=1, G=1,$), which is a very small value.
Note that the shear has four maxima and three minima, with one minimum being exactly zero 
and there are two maxima with a larger value. If we consider initial 
conditions for $p_1,p_2,p_3$ with larger values, 
then the maxima of the shear become smaller than the current solution,
shown in figures \ref{fig:nonshear-vol} and \ref{fig:nonshear-shear}.
Next, we shall add a level of complexity and study the Bianchi I limit in what follows.

\subsection{Bianchi I Limit}
\label{sec:B1-lim}

\begin{figure}[tpb!]
\begin{center}
\includegraphics[width=11cm]{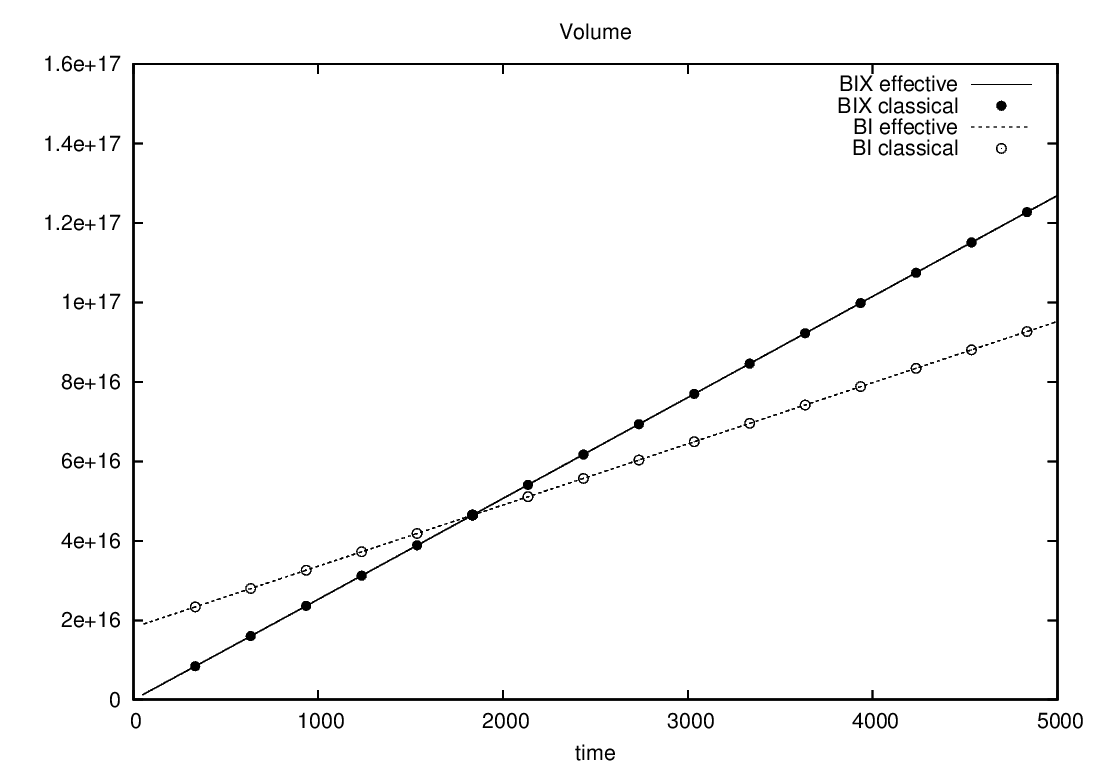}
\caption{Time evolution of the volume for Bianchi IX and Bianchi I, effective and classical.
Note that the solutions coincide at time $t=1835$ where the initial conditions are chosen: 
$\bar\mu_1 c_1 = 6.0\times 10^{-5}$,
$\bar\mu_2 c_2 = -1.1\times 10^{-5}$,
$\bar\mu_3 c_3 = 1.3\times 10^{-4}$,
$p_1 = 6.0\times 10^{10}$,
$p_2 = 4.0\times 10^{11}$,
$p_3 = 9.0\times 10^{10}$. 
}
\label{fig:b9-b1-volume}
\end{center}
\end{figure}

\begin{figure}[tpb!]
\begin{center}
\begin{tabular}{cc}
\includegraphics[width=11cm]{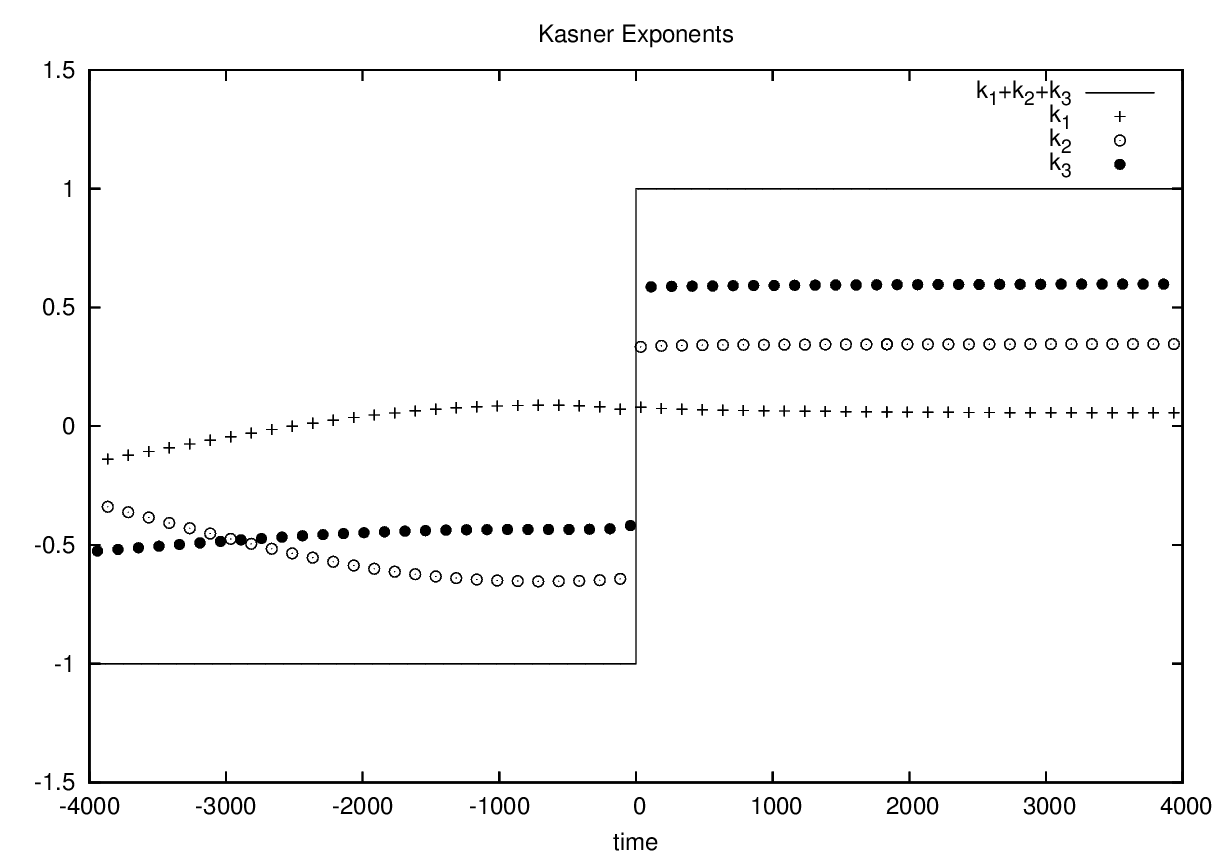} \\
\includegraphics[width=11cm]{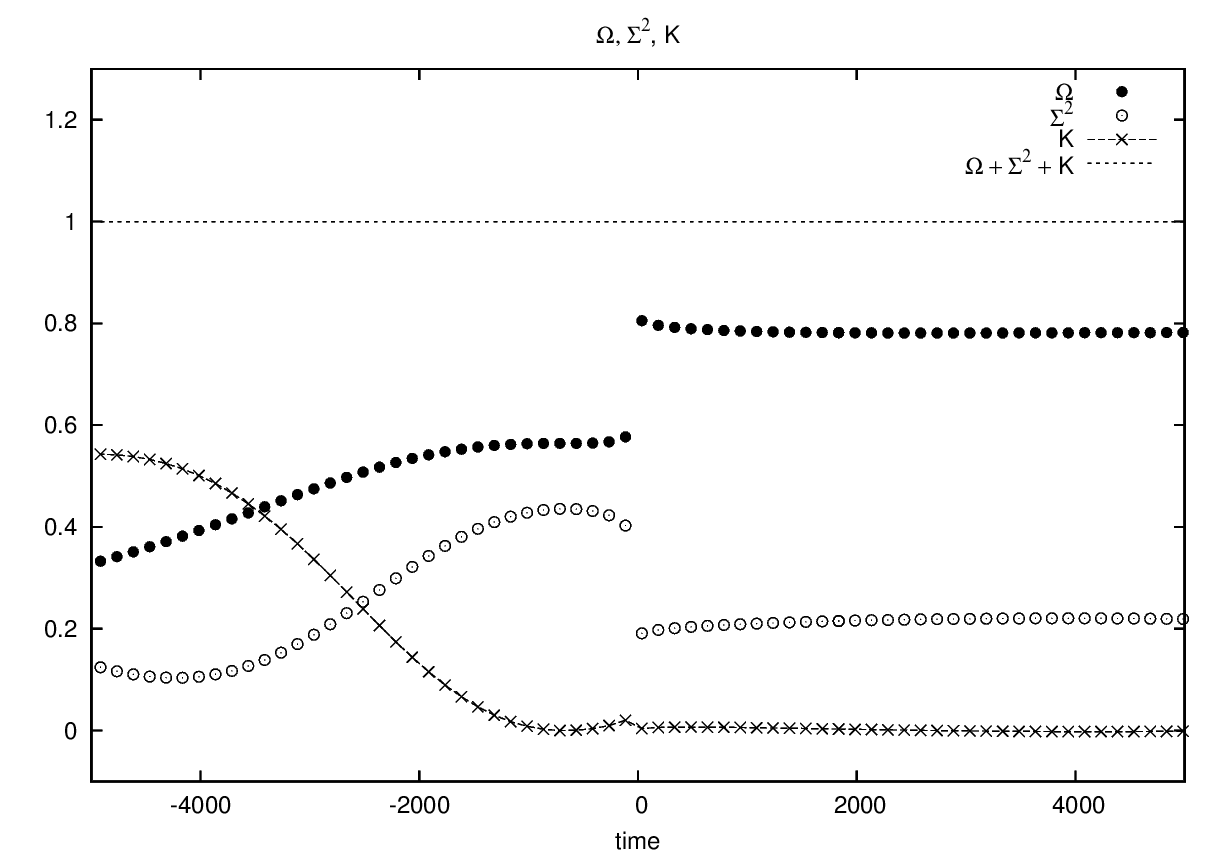}
\end{tabular}
\caption{Evolution of the Kasner exponents ($k_i$) and the parameters that 
measure the dynamical contribution of matter ($\Omega$), anisotropies ($\Sigma^2$) 
and intrinsic curvature ($K$). Recall that these quantities are not well defined at the bounce, so what we
see is a transition from one value on one side of the bounce to a different value at the other side.
The initial conditions at the time $t = 1835$ are:
$\bar\mu_1 c_1 = 6.0\times 10^{-5}$,
$\bar\mu_2 c_2 = -1.1\times 10^{-5}$,
$\bar\mu_3 c_3 = 1.3\times 10^{-4}$,
$p_1 = 6.0\times 10^{10}$,
$p_2 = 4.0\times 10^{11}$,
$p_3 = 9.0\times 10^{10}$. 
}
\label{fig:b9-b1-kasner}
\end{center}
\end{figure}

\begin{figure}[tpb!]
\begin{center}
\begin{tabular}{cc}
\includegraphics[width=0.5 \textwidth]{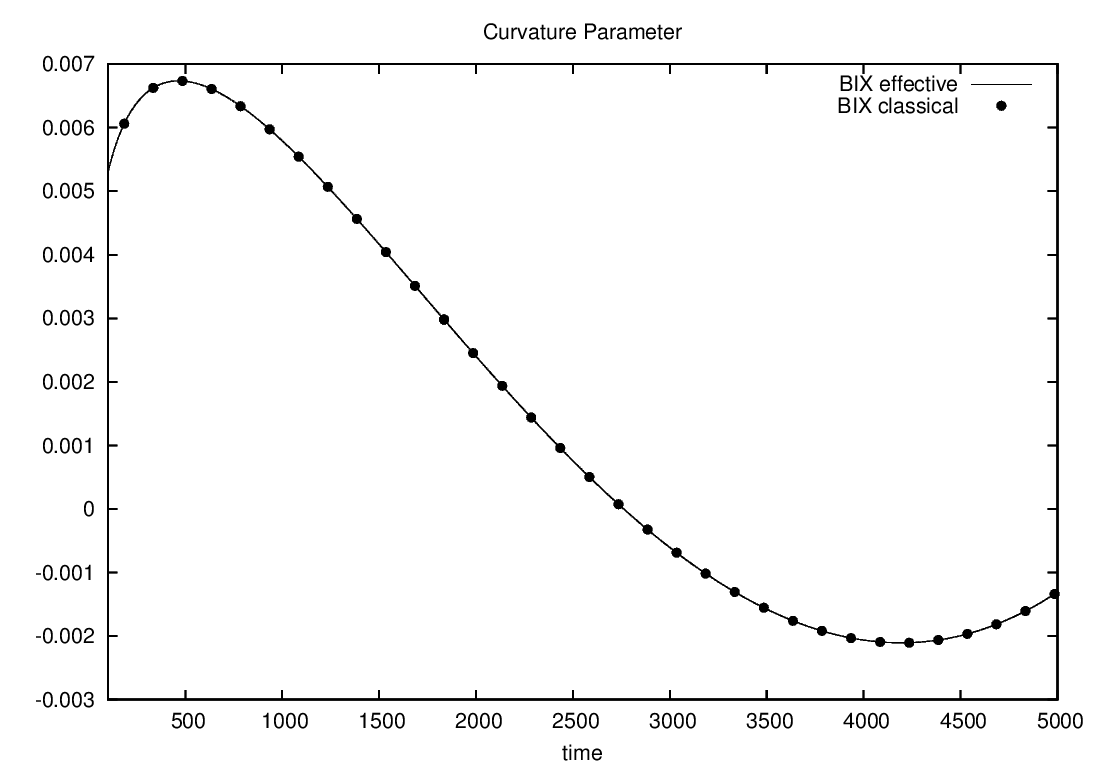}
\includegraphics[width=0.5 \textwidth]{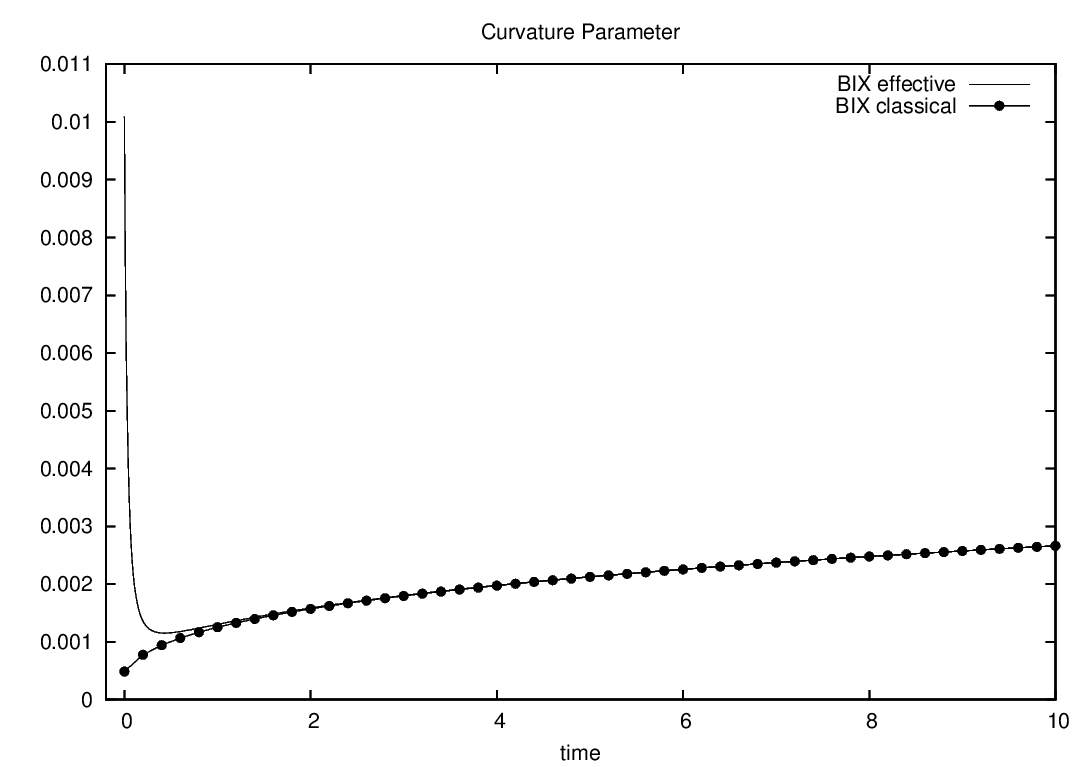}
\end{tabular}
\caption{Time evolution of the curvature parameter ($K$) for the effective and classical Bianchi IX.
Note that $K$ is near to zero but has a non trivial behavior, as it becomes negative even at classical level. In the right figure we make a zoom up until the point where the effective and classical dynamics differ, near the bounce (recall that the quantity $K$ is not well defined at the bounce).
The initial conditions are: 
$\bar\mu_1 c_1 = 6.0\times 10^{-5}$,
$\bar\mu_2 c_2 = -1.1\times 10^{-5}$,
$\bar\mu_3 c_3 = 1.3\times 10^{-4}$,
$p_1 = 6.0\times 10^{10}$,
$p_2 = 4.0\times 10^{11}$,
$p_3 = 9.0\times 10^{10}$. 
}
\label{fig:b9-b1-curvature}
\end{center}
\end{figure}

\begin{figure}[tpb!]
\begin{center}
\begin{tabular}{cc}
\includegraphics[width=0.5 \textwidth]{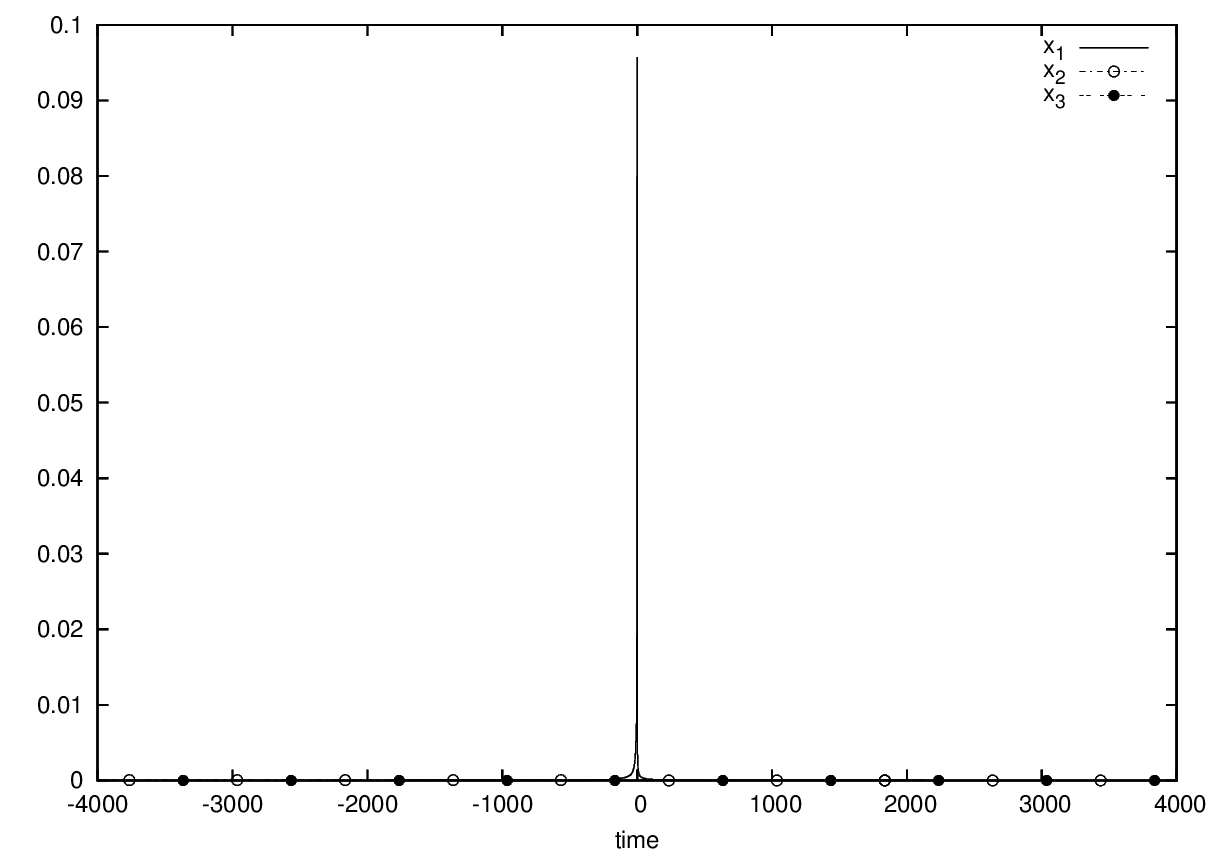} &
\includegraphics[width=0.5 \textwidth]{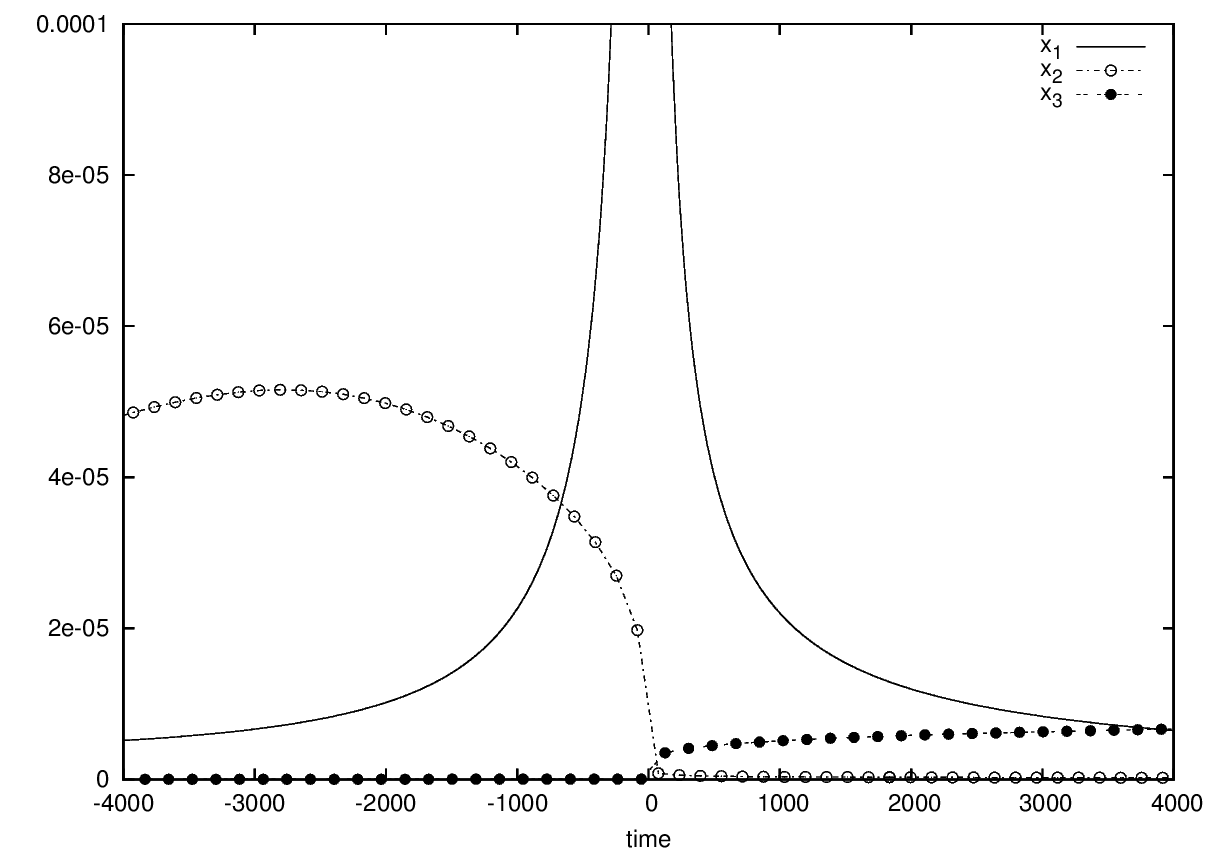}
\end{tabular}
\caption{Evolution of the $x_i$ quantities that contribute to the intrinsic curvature.
The initial conditions at the time $t = 1835$ are:
$\bar\mu_1 c_1 = 6.0\times 10^{-5}$,
$\bar\mu_2 c_2 = -1.1\times 10^{-5}$,
$\bar\mu_3 c_3 = 1.3\times 10^{-4}$,
$p_1 = 6.0\times 10^{10}$,
$p_2 = 4.0\times 10^{11}$,
$p_3 = 9.0\times 10^{10}$. 
}
\label{fig:b9-b1-xi}
\end{center}
\end{figure}

In order to find those Bianchi IX solutions that are close to the Bianchi I solutions
we employ the following strategy:
\bi
\item Impose the initial conditions at the bounce for the effective Bianchi I
equations, such that the field momentum is large.
\item Evolve forward in time the Bianchi I equations until the classical region.
\item Take the values from the Bianchi I evolution and use them as initial
conditions for the Bianchi IX effective equations.
\item Evolve back in time the effective Bianchi IX equations. 
\ei

Using this procedure we get Bianchi IX solutions that start as a Bianchi I.
The initial conditions that come from the Bianchi I evolution are:
$\bar\mu_1 c_1 = 6.0\times 10^{-5}$,
$\bar\mu_2 c_2 = -1.1\times 10^{-5}$,
$\bar\mu_3 c_3 = 1.3\times 10^{-4}$,
$p_1 = 6.0\times 10^{10}$,
$p_2 = 4.0\times 10^{11}$,
$p_3 = 9.0\times 10^{10}$, the momentum of the field is $p_\phi \sim 1.83 \times 10^{12}$. 
The initial time is $t = 1835$ and the Bianchi IX equations are evolved back and forward in time. 
The momentum of the field for the Bianchi IX constraint is $p_\phi \sim 3.66\times 10^{12}$, 
note that it is not exactly equal to the Bianchi I constraint, this is due to fact that Bianchi
IX solutions should not necessarily follow the Bianchi I solutions the entire time, even in General 
Relativity. It is shown in figure \ref{fig:b9-b1-volume}, 
where can be seen that the Bianchi IX and Bianchi I solutions
coincide only at the initial conditions, and both follow the classical behaviour for each corresponding
model.
\medskip

Figure~\ref{fig:b9-b1-kasner} depicts the evolution of the Kasner exponents and the parameters
$\Omega, \Sigma^2, K$, which remain almost constant in the region after the bounce. 
This is what we expect from a solution that evolves as a Bianchi I universe.
The Kasner transitions in Bianchi I were studied in detail in \cite{singh-gupt-2}.
In figure \ref{fig:b9-b1-kasner} it can be seen than the dynamical contribution from 
the intrinsic curvature is near zero 
($K\sim 0$), which is a condition that Bianchi I universes satisfy,
but it has a non trivial behaviour, figure \ref{fig:b9-b1-curvature}. Note that $K$ becomes negative
even at classical level.

The $x_i$ are the quantities that contribute to the intrinsic curvature ${}^{(3)}R$, Eq.~\eqref{Ricci3},
whose evolution is shown in figure \ref{fig:b9-b1-xi}, 
where it can be seen that $x_i\sim 0$ after the bounce, while at the bounce $x_1$ is large,
and the other two are close to zero ($x_2, x_3\sim 0$). Recall that these are precisely the conditions
that define the Bianchi II limit. 
Therefore, at the bounce there is a Bianchi II transition.
The evolution before the bounce
is highly non trivial, the dynamical contribution is changing constantly 
between $\Omega, \Sigma^2$ and $ K$ (Fig.~\ref{fig:b9-b1-kasner}), now $x_1$ and $ x_2$ are contributing to $K$ and 
$x_3$ remains close to zero, which we have classified as the ``Bianchi VII${}_0$ limit''. 
Of course, as we mentioned before, the Bianchi VII${}_0$ is just a label and needs to be 
properly studied to determine whether this dynamics really correspond to the effective 
dynamics of the loop quantum cosmology  Bianchi VII${}_0$ model.
\medskip

To summarize, in Figs. \ref{fig:b9-b1-kasner} and \ref{fig:b9-b1-xi} it
can be appreciated that the evolution has different stages. At 
first the system evolves close to the Bianchi VII${}_0$ limit ($x_3\sim 0$ and $K\ne 0$).
Later on, it has a Bianchi II transition ($x_2, x_3\sim 0$ and $x_1\ne 0$) at the bounce 
and, finally, it evolves like a the Bianchi I ($K\sim 0$, $\Omega, \Sigma^2, k_i$ constants).
This is a full Bianchi IX effective dynamics that can be characterized with some phases and
transitions. Note that one can not assure that all the solutions can be characterized in this way;
indeed it is highly probable that the dynamics in the full space of solutions would have even
richer behaviours. 

Finally, we want to add some comments about the evolution of $\Omega, \Sigma^2, K$ 
in Fig.~\ref{fig:b9-b1-kasner}. First, the discontinuity of $\Omega, \Sigma^2, K$ is 
because they are ill-defined at the bounce (they are defined as a ratio over $\theta$). 
Second, it is not completely clear in the figure, but the curvature parameter $K$ is negative in some
interval of time before the bounce. Third, the evolution before $t=-5000$ could not be performed
due to the accumulation of numerical errors. Thus, it is not possible to try to reach any conclusions
without a better numerical integration. This is due to the equations of motion being stiff
and needing a special treatment for long time evolutions.

\section{Conclusions}
\label{sec:5}

In this manuscript, we studied the numerical solutions to the effective Bianchi IX
equations in loop quantum cosmology. This allowed us to answer some of the questions that motivated
this work. Let us start by summarizing our results. Throughout this work, we considered two sets
of effective equations, differing in their treatment of inverse triad corrections.
We started with the question regarding the resolution of the big bang singularity. 
In sections \ref{sec:big-vol} and \ref{sec:lim-clas} we showed that both effective theories
in the large field momentum limit describe the same dynamics, 
resolve the singularity, and replace it by a bounce, 
and far from the bounce reproduce the classical dynamics. Furthermore, given that the Bianchi IX 
model has positive spatial curvature, 
then there is a infinite number of bounces and recollapses. \medskip

We found in sections \ref{sec:iso-lim} and \ref{sec:B1-lim} 
that the set of effective Bianchi IX has as a limit the effective Bianchi I, and the effective 
isotropic FLRW with $k=0$.
The FLRW with $k=1$ is contained within the Bianchi IX model only if the inverse triad corrections 
are not included (for the case with lapse $N$=$V$). When they are included (for lapse $N$=1), 
the FLRW is not embedded in Bianchi IX, but it is rather a limiting  case when the volume at
the bounce is large. 
\medskip

Additionally, in section \ref{sec:B1-lim} we showed that the effective dynamics of Bianchi IX can be
characterized by different phases. These phases can be classified according to the values of the
parameters $x_i$ introduced in section \ref{sec:observables}, which contribute to the 
intrinsic curvature. Moreover, the Kasner parameters $k_i$ are useful 
to characterize the Bianchi IX dynamics. In our case, the universe evolves from a 
universe close to Bianchi VII${}_0$ 
($x_3\sim 0$), later on it undergoes a bounce in which it exhibits a  Bianchi II type 
transition ($x_2, x_3\sim 0$), 
and finally evolves close to a Bianchi I solution ($x_1, x_2, x_3\sim 0$ and $k_1,k_2,k_3$ constants)
far from the bounce.
Note that this is the first time that the Bianchi VII${}_0$ model is mentioned in the context of loop
quantum cosmology. This model has not been studied yet, but we are speculating that the solutions 
that we get are close to what would be the effective dynamics of Bianchi VII${}_0$, which would come
from the ``improved dynamics'' Bianchi VII${}_0$ loop quantization.
\medskip

All the result that we have obtained are in the limit of large field momentum, where both effective 
theories of Bianchi IX have the same dynamics. If these two theories really describe the quantum 
dynamics of the semiclassical states, as it is expected, then our results indicate that the 
semiclassical dynamics of both quantum theories 
describe the same physics for a realistic large universe.
\medskip

We shall end by discussing some related issues to what we have studied in this manuscript.
\begin{enumerate}
\item When one considers many bounces and recollapses, then new features and interesting behaviours appear,
when looking at the dynamics of the observables \cite{CKM}, especially in the vacuum case that is probably
the most interesting one. 
\item One can also perform qualitative studies that are independent of
the matter content of the theory, by just imposing energy conditions. 
In that case, one can explore what happens to the BKL behaviour when quantum corrections
--as captured by the effective dynamics-- are incorporated. Those rather intriguing results
will be published in a forthcoming communication \cite{CKM-short}.
\item In this manuscript we have considered only two effective theories of Bianchi IX, 
but there are more effective theories \cite{singh-gupt,singh-Ed}
which explore the ambiguities in the quantization of Bianchi IX. 
\item Preliminary studies \cite{CKM-short} indicate that our results will also apply to the 
other effective theories \cite{singh-gupt,singh-Ed} in the limit of large field momentum.
\end{enumerate}

\section*{Acknowledgments}
We thank A. Karami and B. Gupt for helpful discussions and comments, and the referees, whose comments have improved the manuscript.
This work was in part supported by DGAPA-UNAM IN103610 grant, by CONACyT 0177840 
and 0232902 grants, by the PASPA-DGAPA program, by NSF PHY-1505411 and
PHY-1403943 grants, by the Eberly Research Funds of Penn State 
and by Patrimonio Aut\'onomo fondo nacional de financiamiento para la ciencia, la tecnolog\'ia 
y la innovaci\'on, Francisco Jos\'e de Caldas.

\begin{appendix}
\section{Effective Equations of Motion}
\label{app:b3}

In this section we collect all the effective equations of motion that we use
in the numerical study of the solutions.
Some of these equations of motion come directly from the original articles,
we just want to put all together in this appendix. The programs developed 
are available upon request.

\subsection*{Flat FLRW, $k=0$}

The effective Hamiltonian is \cite{aps,aps2}
\begin{equation}
\mc{C}_{\rm H_{k=0}}=\frac{3}{8\pi G\gamma^2\lambda^2}\,V^2 \,\sin^2(\lambda\beta) -\frac{p_\phi^2}{2}\approx 0\, ,
\end{equation}

where the lapse is $N=V$, with $V$ the physical volume of the fiducial cell.
The Poisson brackets are $\{\beta,V\}=4\pi G \gamma$ and $\{\phi,p_\phi\}=1$. 
The variables $(V,\beta)$ are related with $(c,p)$ by
\be 
V=p^{3/2}\,,\quad\beta = \f{c}{ \sqrt{p} }\,.
\ee 

The equations of motion for $(V,\beta)$ are 
\ba
\dot{V} &=& \frac{3}{\gamma\lambda}\,V^2 \,\sin(\lambda\beta)\,\cos(\lambda\beta)\,,\\
\dot{\beta} &=& -\frac{3}{\gamma\lambda^2}\,V \,\sin^2(\lambda\beta)\,.
\ea

\subsection*{Closed FLRW, $k=1$}

For the closed FLRW model there are different quantum theories that depend
on the quantization method used for defining the curvature \cite{CK}, and also depending on whether the inverse triads corrections are included or not.
For our study we choose the equations that come from the quantization where the curvature is calculated using 
the connection, which is obtained  from the `open' holonomies. Even so the 
freedom to choose the inverse triad correction remains. Therefore we are 
going to consider two effective theories, depending of the lapse function.
The effective Hamiltonian with lapse $N=V$ is given by \cite{CK,Asieh}
\ba
\mc{C}_{\rm H_{k=1}}^{(1)}&=&\frac{3V^2}{8\pi G\gamma^2 \lambda^2}
\big[\sin^2\lambda\beta-2D\sin\lambda\beta+(1+\gamma^2)D^2\big]-\f{p_\phi^2}{2}\approx 0 \, .
\ea 
The equations of motion are
\ba
\dot{V}&=&\frac{3}{\gamma\lambda}\,V\,\cos(\lambda\beta)\,[\sin(\lambda\beta)-D]\,,\\
\dot{\beta} &=& -\frac{1}{2\gamma \lambda^2}
\big[ 3\sin^2\lambda\beta-4D\sin\lambda\beta+(1+\gamma^2)D^2\big]
-2\pi G\gamma\f{p_\phi^2}{V^2}\,.
\ea

The effective Hamiltonian with lapse $N=1$ is \cite{CK,Asieh}
\ba
\mc{C}_{\rm H_{k=1}}^{(2)}&=&\frac{3A(V)V}{8\pi G\gamma^2 \lambda^2}
\big[\sin^2\lambda\beta-2D\sin\lambda\beta+(1+\gamma^2)D^2\big]-\f{p_\phi^2}{2}\approx 0  \, ,
\ea 
with 
\be
A(V)=\frac{1}{2V_c}(V+V_c-|V-V_c|)=
\left\{
\begin{array}{lr}
V/V_c &,\,{\rm if }\, V< V_c\\
1 &, \,{\rm if }\, V\geq V_c
\end{array}
\right.
\ee 

and $V_c=2\pi\gamma\lambda\lp^2$. 
The equations of motion are
\ba 
V'&=&\frac{3}{\gamma\lambda}A(V)\cos\lambda\beta(V\sin\lambda\beta-\lambda\vartheta V^{2/3})\,,\\
\beta'&=&-\frac{3}{2\gamma\lambda^2}
\Bigg[\bigg(VA_{,V}+A(V)\bigg)\sin^2\lambda\beta-2\lambda\vartheta\sin\lambda\beta
\bigg(\frac{2}{3}A(V)V^{-1/3}+A_{,V}V^{2/3}\bigg)\nonumber\\
& &+\lambda^2\vartheta^2(1+\gamma^2)
\bigg(\frac{A(V)V^{-2/3}}{3}+A_{,V}V^{1/3}\bigg)\Bigg]
-2\pi G\gamma\frac{p_\phi^2}{V^2}\,,
\ea  
where
\be
A_{,V}=\frac{1}{2V_c}\left(1-\frac{|V-V_c|}{V-V_c}\right)
=\left\{
\begin{array}{lr} 
1/V_c &,\,{\rm if }\,  V<V_c\\ 
0 & ,\,{\rm if }\, V>V_c
\end{array}
\right.
\ee

In both theories the definition of the $(V,\beta)$ variables is the same
than the flat FLRW case ($V=p^{3/2}, \beta = {c}/{ \sqrt{p} }, \{\beta,V\}=4\pi G \gamma$),
but now the volume correspond to the physical volume of all the space.

\subsection*{Bianchi I and II}
The effective Hamiltonians for Bianchi I and II can be written in one single
expression \cite{bianchiII}
\ba
\mc{C}_{\rm H_{\rm BII}} & =& \f{p_1p_2p_3}{8\pi G\gamma^2\lambda^2}
\Big[ \sin\bar\mu_1c_1\sin\bar\mu_2c_2
+\sin\bar\mu_2c_2\sin\bar\mu_3c_3
+\sin\bar\mu_3c_3\sin\bar\mu_1c_1 \Big] \nonumber\\
& &\quad + \f{1}{8\pi G\gamma^2}
\Bigg[\f{\alpha(p_2p_3)^{3/2}}{\lambda\sqrt{p_1}}\sin\bar\mu_1c_1
-(1+\gamma^2)\left(\f{\alpha p_2p_3}{2p_1}\right)^2 \Bigg] - \f{p_\phi^2}{2} \approx 0 \, ,
\ea 
where $\alpha$ allows to choose between Bianchi I ($\alpha=0$) and Bianchi II ($\alpha = 1$).
The Poisson brackets are $\{c^i,p_j\}=8\pi G\gamma\delta_j^i$ and $\{\phi,p_\phi\}=1$.
The equations of motion are \cite{bianchiII}
\ba
\dot{p_1} &=& \f{p_1^2}{\gamma\bar\mu_1}\left(\sin\bar\mu_2c_2+
\sin\bar\mu_3c_3+\lambda x \right)\cos\bar\mu_1c_1, \\
\dot{p_2} &=& \f{p_2^2}{\gamma\bar\mu_2}(\sin\bar\mu_1c_1+\sin\bar\mu_3c_3)
\cos\bar\mu_2c_2, \\
\dot{p_3} &=& \f{p_3^2}{\gamma\bar\mu_3}(\sin\bar\mu_1c_1+\sin\bar\mu_2c_2)
\cos\bar\mu_3c_3, 
\ea 
\ba
\dot{c_1} &=& -\f{p_2p_3}{2\gamma\lambda^2}\Big[
2(\sin\bar\mu_1c_1\sin\bar\mu_2c_2+\sin\bar\mu_1c_1\sin\bar\mu_3c_3
+\sin\bar\mu_2c_2\sin\bar\mu_3c_3) \nonumber \\ 
& &\qquad +\bar\mu_1c_1\cos\bar\mu_1c_1
(\sin\bar\mu_2c_2+\sin\bar\mu_3c_3)-\bar\mu_2c_2\cos\bar\mu_2c_2
(\sin\bar\mu_1c_1+\sin\bar\mu_3c_3) \nonumber  \\ 
& &\qquad -\bar\mu_3c_3\cos\bar\mu_3c_3
(\sin\bar\mu_1c_1+\sin\bar\mu_2c_2)+{\lambda^2x^2}(1+\gamma^2) \nonumber \\ 
& &\qquad +{\lambda x}(\bar\mu_1c_1\cos\bar\mu_1c_1-\sin\bar\mu_1c_1) \Big], 
\ea 
\ba 
\dot{c_2} &=& -\f{p_1p_3}{2\gamma\lambda^2}\Big[
2(\sin\bar\mu_1c_1\sin\bar\mu_2c_2+\sin\bar\mu_1c_1\sin\bar\mu_3c_3
+\sin\bar\mu_2c_2\sin\bar\mu_3c_3) \nonumber \\ 
& &\qquad -\bar\mu_1c_1\cos\bar\mu_1c_1
(\sin\bar\mu_2c_2+\sin\bar\mu_3c_3)+\bar\mu_2c_2\cos\bar\mu_2c_2
(\sin\bar\mu_1c_1+\sin\bar\mu_3c_3) \nonumber \\ 
& &\left.\qquad -\bar\mu_3c_3
\cos\bar\mu_3c_3(\sin\bar\mu_1c_1+\sin\bar\mu_2c_2)\right]
-{\lambda^2x^2}(1+\gamma^2) \nonumber \\ 
& &\qquad -{\lambda x}(\bar\mu_1c_1\cos\bar\mu_1c_1 -3\sin\bar\mu_1c_1) \Big], 
\ea
\ba 
\dot{c_3} &=& -\f{p_1p_2}{2\gamma\lambda^2}\Big[ 
2( \sin\bar\mu_1c_1\sin\bar\mu_2c_2+\sin\bar\mu_1c_1\sin\bar\mu_3c_3
+\sin\bar\mu_2c_2\sin\bar\mu_3c_3) \nonumber \\ 
& &\qquad -\bar\mu_1c_1\cos\bar\mu_1c_1
(\sin\bar\mu_2c_2+\sin\bar\mu_3c_3)-\bar\mu_2c_2\cos\bar\mu_2c_2
(\sin\bar\mu_1c_1+\sin\bar\mu_3c_3) \nonumber \\ 
& &\left.\qquad +\bar\mu_3c_3
\cos\bar\mu_3c_3(\sin\bar\mu_1c_1+\sin\bar\mu_2c_2)\right]
-{\lambda^2x^2}(1+\gamma^2) \nonumber \\ 
& & \qquad -{\lambda x}(\bar\mu_1c_1\cos\bar\mu_1c_1 -3\sin\bar\mu_1c_1) \Big],
\ea 

with 
\be 
x=\alpha\sqrt{\frac{p_2 p_3}{p_1^3}}\,.
\ee

\subsection*{Bianchi IX}

The effective Hamiltonian for Bianchi IX with lapse $N=V$ is \cite{bianchiIX}
\ba  
\mc{C}_{\rm H_{\rm BIX}}^{(1)} &=&
-\f{p_1p_2p_3}{8\pi G\gamma^2\lambda^2}
\big(\sin\bar\mu_1c_1\sin\bar\mu_2c_2+\sin\bar\mu_2c_2
\sin\bar\mu_3c_3+\sin\bar\mu_3c_3\sin\bar\mu_1c_1\big) \nonumber\\ 
& & -\f{\vartheta}{4\pi G\gamma^2\lambda}\bigg(
\f{(p_1p_2)^{3/2}}{\sqrt{p_3}}\sin\bar\mu_3c_3+
\f{(p_2p_3)^{3/2}}{\sqrt{p_1}}\sin\bar\mu_1c_1+
\f{(p_3p_1)^{3/2}}{\sqrt{p_2}}\sin\bar\mu_2c_2\bigg)  \nonumber\\
& & -\f{\vartheta^2 (1+\gamma^2)}{8\pi G \gamma^2}
\bigg[2(p_1^2+p_2^2+p_3^2) -
\left(\f{p_1p_2}{p_3}\right)^2
-\left(\f{p_2p_3}{p_1}\right)^2
-\left(\f{p_3p_1}{p_2}\right)^2
\bigg] \nonumber \\
& & + \f{p_\phi^2}{2} \approx 0 \,. 
\ea 

The equations of motion are \cite{bianchiIX}
\ba
\dot{p_1} &=&
\frac{1}{\gamma}\left[\f{p_1^2}{\bar\mu_1}(\sin\bar\mu_2c_2+
\sin\bar\mu_3c_3)+2\vartheta p_2p_3\right]\cos\bar\mu_1c_1\,,\\
\dot{c_1} &=& -\f{1}{\gamma}\left\{ \f{p_2p_3}{\lambda^2}\bigg(
\sin\bar\mu_1c_1\sin\bar\mu_2c_2+\sin\bar\mu_1c_1\sin\bar\mu_3c_3+\sin\bar\mu_2
c_2\sin\bar\mu_3c_3 \right. \nonumber\\ 
& & \qquad +\f{\bar\mu_1c_1}{2}\cos\bar\mu_1c_1(
\sin\bar\mu_2c_2+\sin\bar\mu_3c_3)-\f{\bar\mu_2c_2}{2}\cos\bar\mu_2c_2(\sin\bar
\mu_1c_1+\sin\bar\mu_3c_3)  \nonumber\\ 
& & \qquad -\f{\bar\mu_3c_3}{2}\cos\bar
\mu_3c_3(\sin\bar\mu_1c_1+\sin\bar\mu_2c_2)\bigg)
+2\vartheta\bigg(\f{3}{2\bar\mu_1}\bigg[
\f{p_1p_2}{p_3}\sin\bar\mu_3c_3
+\f{p_1p_3}{p_2}\sin\bar\mu_2c_2  \nonumber\\ 
& & \qquad
-\f{p_2p_3}{3p_1}\sin\bar\mu_1c_1\bigg]
+\f{1}{2}\f{p_2p_3}{p_1}c_1\cos\bar\mu_1c_1
-\f{1}{2}p_2c_3\cos\bar\mu_3c_3
-\f{1}{2}p_3c_2\cos\bar\mu_2c_2\bigg)  \nonumber\\ 
& &\qquad \left. 
+{\vartheta^2}(1+\gamma^2)\bigg[4p_1-2p_1\left(\f{p_2^2}{p_3^2}
+\f{p_3^2}{p_2^2}\right)+2\f{p_2^2p_3^2}{p_1^3}
\bigg]\right\}\,. 
\ea 

The other equations of motion are obtain by permutations. 
The effective Hamiltonian for Bianchi IX with lapse $N=1$ is given by \cite{CK-3,Asieh}
\ba 
\mc{C}_{\rm H_{BIX}}^{(2)}&=&
-\frac{V^4 A(V)h^6(V)}{8\pi G\gamma^2\lambda^2 V_c^6}
\big(\sin\bar\mu_1c_1\sin\bar\mu_2c_2 +\sin\bar\mu_1c_1\sin\bar\mu_3c_3
+\sin\bar\mu_2c_2\sin\bar\mu_3c_3\big) \nonumber\\
& &-\frac{\vartheta A(V)h^4(V)}{4\pi G\gamma^2\lambda V_c^4}\bigg(p_1^2p_2^2\sin\bar\mu_3c_3
+p_2^2p_3^2\sin\bar\mu_1c_1 +p_1^2p_3^2\sin\bar\mu_2c_2\bigg)\nonumber\\
& &-\frac{\vartheta^2(1+\gamma^2)A(V)h^4(V)}{8\pi G\gamma^2 V_c^4} \nonumber\\
& &\quad\quad \times \bigg(2V\big(p_1^2+p_2^2+p_3^2\big) -\big[(p_1p_2)^{4}+(p_1p_3)^{4}+(p_2p_3)^{4}\big]
\frac{h^6(V)}{V_c^6}\bigg)\nonumber\\
& &+\f{h^6(V)V^2}{2V^6_c}p_\phi^2 \approx 0 \,,
\ea
with 
\be 
h(V)=\sqrt{V+V_c}-\sqrt{|V-V_c|}\,.
\ee
The derivatives of $h(V)$ and $A(V)$ are given by
\be
A_{,p_i}=\left\{
\begin{array}{lr}
\frac{1}{V_c}\sqrt{\frac{p_jp_k}{p_i}} &,\,{\rm if }\,  V<V_c\\ 
0 & ,\,{\rm if }\, V>V_c 
\end{array} 
\right. 
\textrm{   ,   } i\neq j\neq k\neq i
\ee 
and
\be
h_{,p_i}=\frac{1}{4}\sqrt{\f{p_jp_k}{p_i}}
\left[\f{1}{\sqrt{V+V_c}}-\frac{\sqrt{|V-V_c|}}{V-V_c}\right]
\textrm{   ,   } i\neq j\neq k\neq i
\ee

The equations of motion are \cite{CK-3,Asieh}
\be
p_1'=\frac{1}{\gamma V_c^4}A(V)V h^4(V)\cos\bar\mu_1c_1
\bigg[\frac{V^2h^2(V)p_1}{V_c^2\lambda}(\sin\bar\mu_2c_2+\sin\bar\mu_3c_3)
-2\vartheta p_2p_3\bigg]\,,
\ee
\be
\begin{split}
c_1'=
&-\frac{h^5(V)}{\gamma\lambda^{2} V_c^6}\bigg(2p_2^2 p_3^2 p_1 A(V) h(V) + V^4 A_{,p_1}h(V) +6V^4 A(V)h_{,p_1}\bigg)\\
&\times\big(\sin\bar\mu_1c_1\sin\bar\mu_2c_2+\sin\bar\mu_1c_1\sin\bar\mu_3c_3
+\sin\bar\mu_2c_2\sin\bar\mu_3c_3\big)\\
&+\frac{2\vartheta}{V_c^4\gamma\lambda}h^3(V)
\bigg[\big(2p_1p_2^2A(V)h(V)+p_1^2p_2^2A_{,p_1}h(V)+4p_1^2p_2^2A(V)h_{,p_1}\big)\sin\bar\mu_3c_3\\
&+\big(2p_1p_3^2A(V)h(V)+p_1^2p_3^2A_{,p_1}h(V)+4p_1^2p_3^2A(V)h_{,p_1}\big)\sin\bar\mu_2c_2\\
&+\big(p_2^2p_3^2A_{,p_1}h(V)+4p_2^2p_3^2A(V)h_{,p_1}\big)\sin\bar\mu_1c_1\bigg]\\
&-A(V)h^4(V)\frac{c_1\cos\bar\mu_1c_1}{2V_c^4\gamma}\left(\frac{V^3h^2(V)}{V_c^2\lambda}
(\sin\bar\mu_2c_2+\sin\bar\mu_3c_3)-2\vartheta\frac{p_2^{3/2}p_3^{3/2}}{p_1^{1/2}}\right)\\
&+A(V)h^4(V)\frac{c_2\cos\bar\mu_2c_2}{2V_c^4\gamma}\left(\frac{p_2^{5/2}p_3^{3/2}p_1^{1/2}h^2(V)}{V_c^2\lambda}
(\sin\bar\mu_1c_1+\sin\bar\mu_3c_3)-2\vartheta p_3V\right)\\
&+A(V)h^4(V)\frac{c_3\cos\bar\mu_3c_3}{2V_c^4\gamma}\left(\frac{p_3^{5/2}p_2^{3/2}p_1^{1/2}h^2(V)}{V_c^2\lambda}
(\sin\bar\mu_1c_1+\sin\bar\mu_2c_2)-2\vartheta p_2V\right)\\
&-\frac{\vartheta^2(1+\gamma^2)}{V_c^4 \gamma} h^3(V)
\Bigg[4 p_1 A(V)V h(V) -\frac{4}{V_c^6}p_1^{3}h^6(V)A(V)(p_2^{4}+p_3^{4})\\
&+(p_1^2+p_2^2+p_3^2)\left(\sqrt{\frac{p_2p_3}{p_1}}A(V)h(V)+8A(V)Vh_{,p_1}+2A_{,p_1}Vh(V)\right)\\
&-\frac{1}{V_c^6}\bigg(10 h^6(V) h_{,p_1} A(V) +h^7(V) A_{,p_1}\bigg)
\bigg(p_1^{4}p_2^{4} +p_1^{4}p_3^{4} +p_2^{4}p_3^{4}\bigg)\Bigg]
-\f{\pi G\gamma p_\phi^2}{p_1\sqrt{p_1p_2p_3}} \,.
\end{split}
\ee

The other equations of motion are obtain by permutations.

\section{Convergence and Conservation}
\label{app:a}

\begin{figure}[ht]
\begin{center}
\begin{tabular}{ll}
\includegraphics[width=0.5 \textwidth]{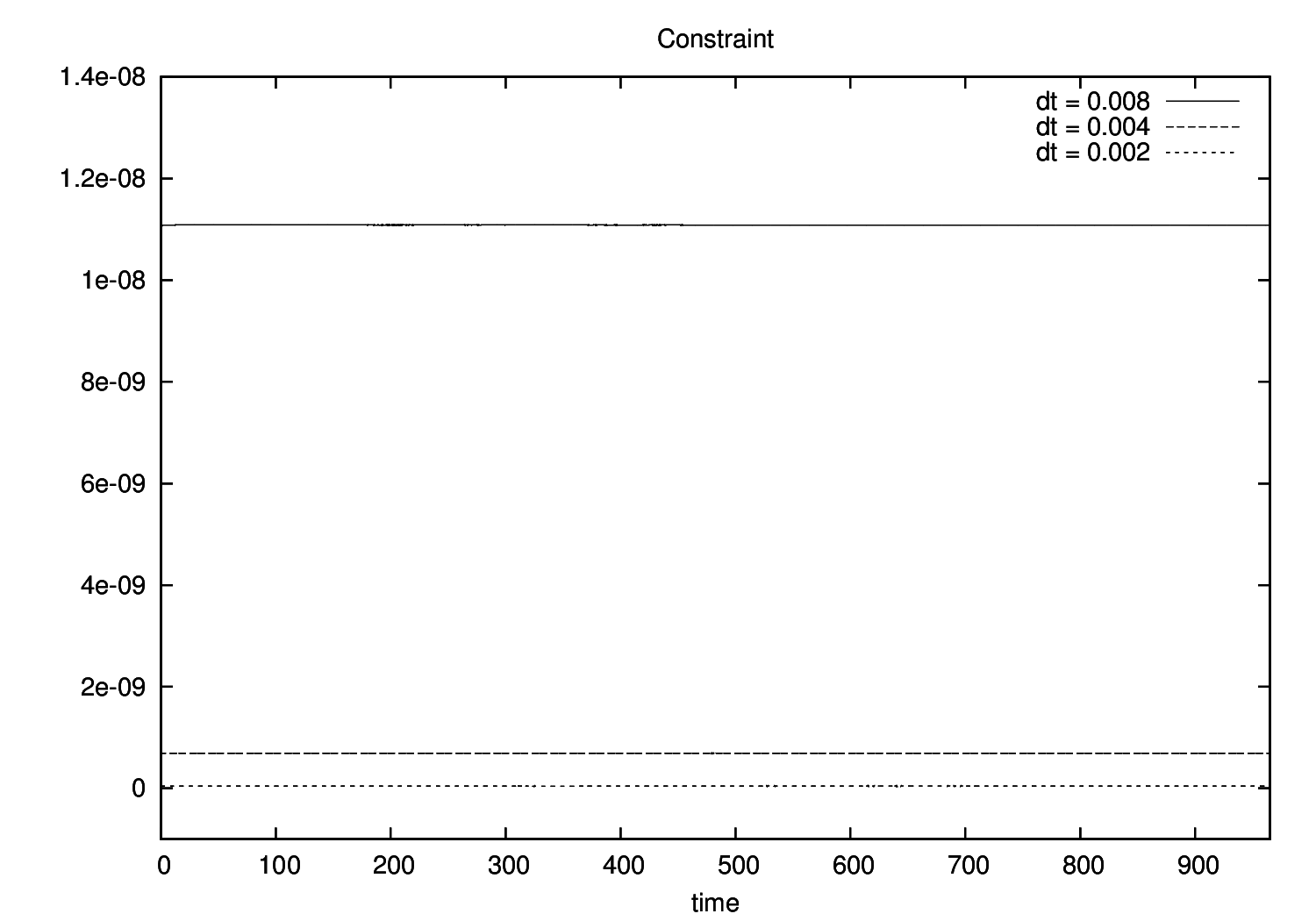}&
\includegraphics[width=0.5 \textwidth]{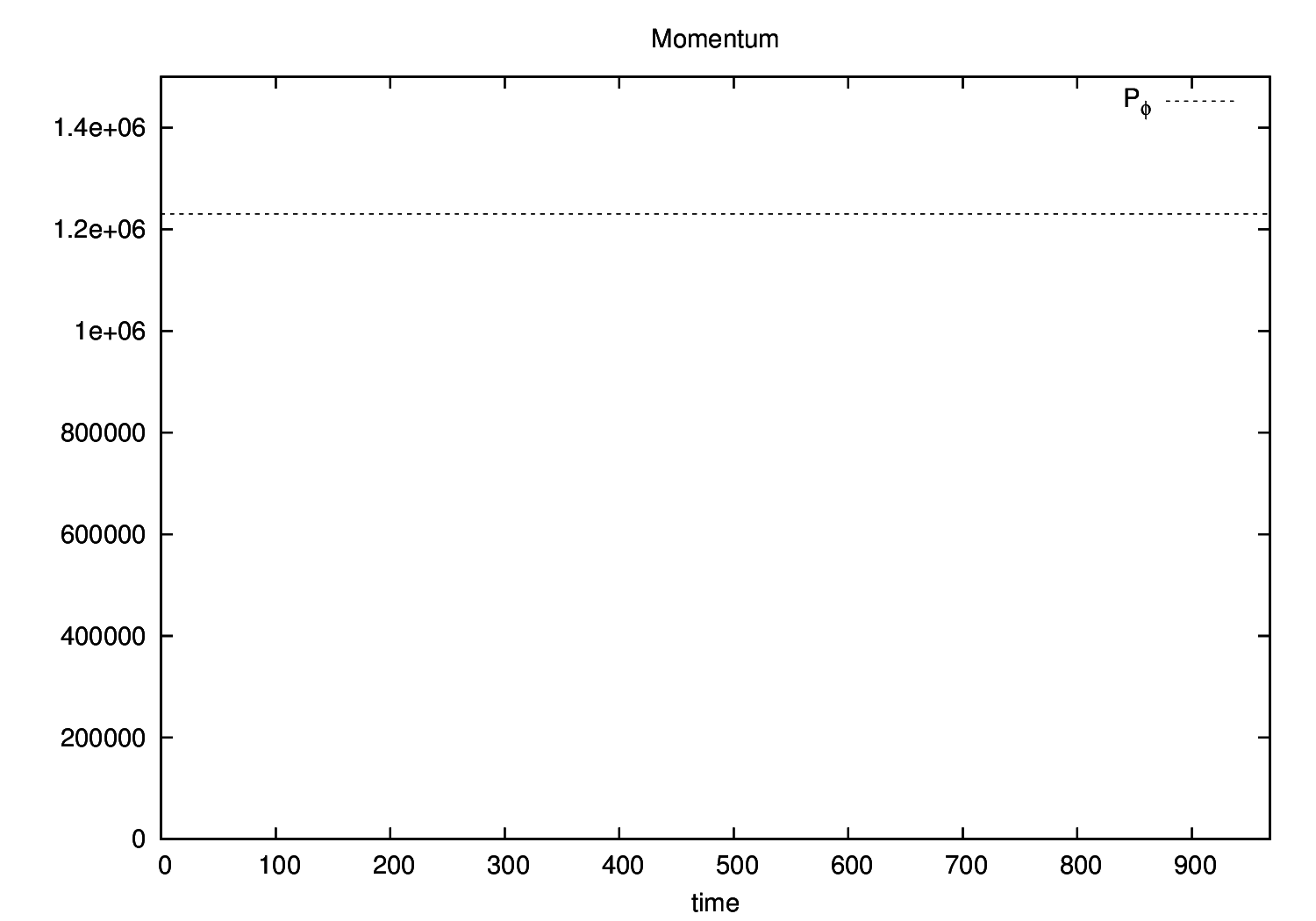}
\end{tabular}
\caption{Convergence of the Hamiltonian constraint and field momentum conservation.
The initial conditions are: 
$\bar\mu_1 c_1= \pi/6$, $\bar\mu_2 c_2= 2\pi/5$, $\bar\mu_3 c_3=5\pi/8$, $p_1=12000$, $p_2=16000$, $p_3=14000$.}
\label{fig:constraint}
\end{center}
\end{figure}

\begin{figure}[htbp!]
\begin{center}
\begin{tabular}{ll}
\includegraphics[width=0.5 \textwidth]{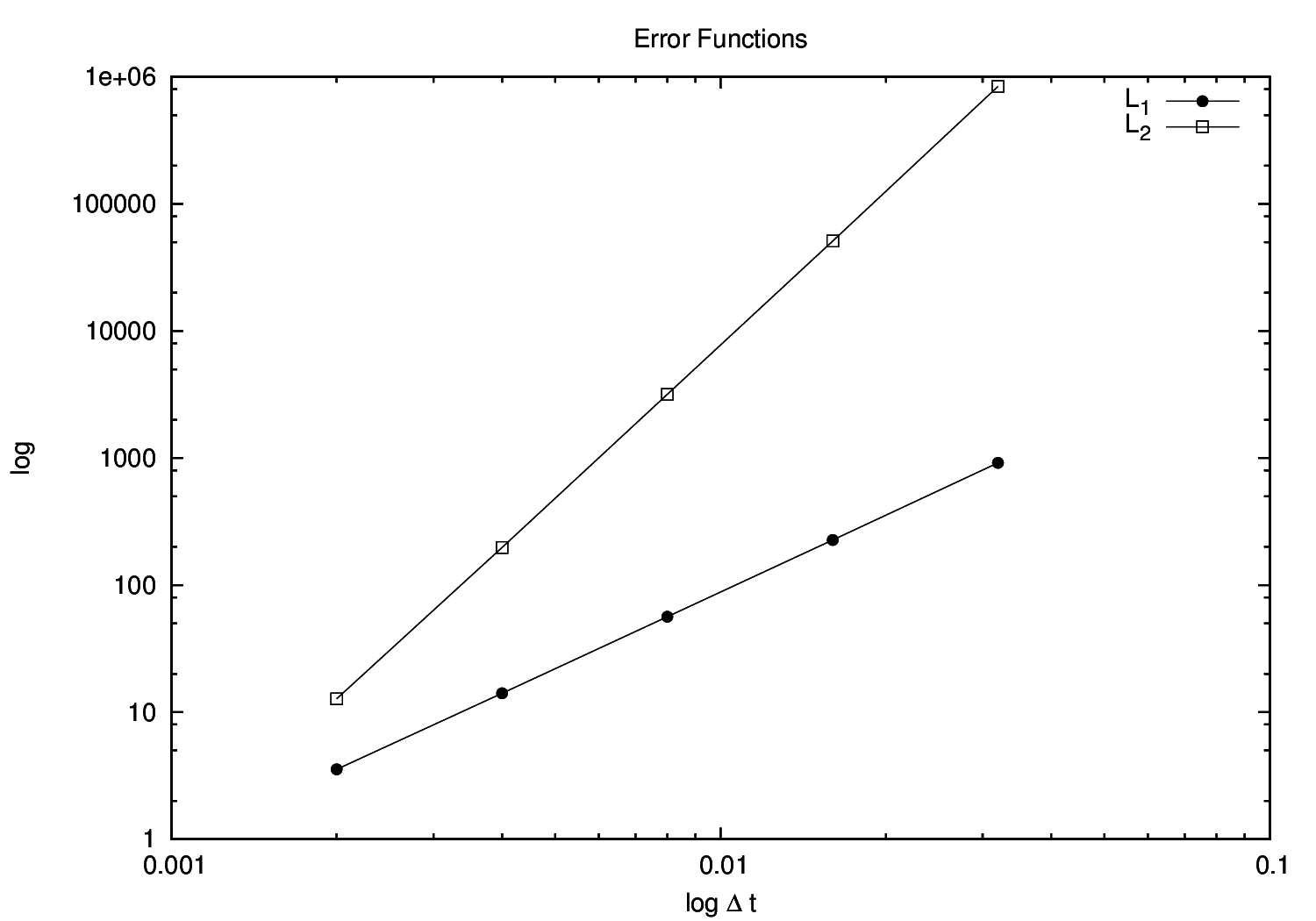}&
\includegraphics[width=0.5 \textwidth]{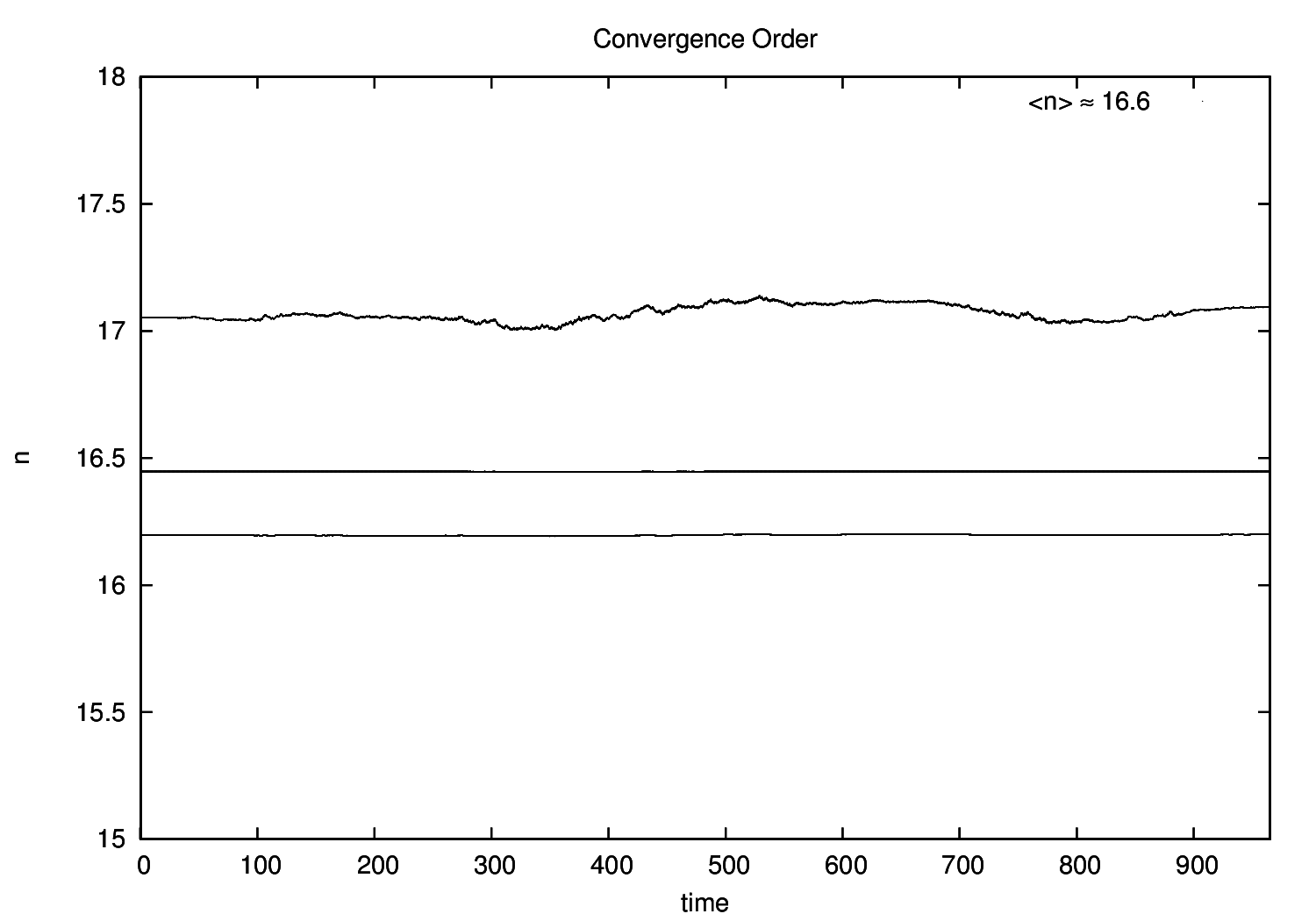}
\end{tabular}
\caption{The first plot shows the error functions $L_1$ and $L_2$. The second plot shows the
convergence order for three different samples. The initial conditions are: 
$\bar\mu_1 c_1= \pi/6$, $\bar\mu_2 c_2= 2\pi/5$, $\bar\mu_3 c_3=5\pi/8$, $p_1=12000$, $p_2=16000$, $p_3=14000$.}
\label{fig:convergence}
\end{center}
\end{figure}

There are two important points that need to be addressed in a numerical work: 
one is the convergence of solutions and the other is
the evolution of conserved quantities. Numerical solutions must also evolve on the constraint surface and they
must preserve the conserved quantities.  This ensures that they are evolving on the physical phase space. 
In figure~\ref{fig:constraint} we show the convergence 
of the Hamiltonian constraint ($\mc{C}_H\approx 0$) and the field momentum evolution, which shows that $p_\phi$ is conserved. The quantities plotted in Figure~\ref{fig:convergence} are the error functions 
\be
L_1 = {\rm Max} |\mc{C}_H(t)_{\rm 2}- \mc{C}_H(t)_{\rm 1}|\, , \quad
L_2 = {\rm Max} \sqrt{|\mc{C}_H^2(t)_{\rm 2}- \mc{C}_H^2(t)_{\rm 1}|}\, ,
\ee
where the sub-indexes in the Hamiltonian constraint mean resolution $2$ (with $\d t_2$) and resolution $1$ 
 (with $\d t_1$),
such that $\d t_2=\d t_1/2$. The method used to integrate the equations is a Runge-Kutta 4 (RK4), 
while the resolutions used for the convergence tests are 
$dt = 0.032, 0.016, 0.008, 0.004, 0.002, 0.001$. We can define the convergence order as
\be
n(t)=\f{f_1-f_2}{f_2-f_3},
\ee
with $f_i$ any evolved function at resolution $i$, with $dt_i>dt_{i+1}$. 
The convergence factor $n$ for a RK4 must be $n=2^4=16$ and we get in our solutions $\langle n \ra \approx 16.6$, 
where $\la n \ra $ denotes the average in time and taking the mean of different samples. 

\end{appendix}


\end{document}